\documentclass[12pt]{iopart}

\usepackage{amssymb}
\usepackage{graphicx}
\usepackage{epstopdf}
\usepackage{bbold}

\usepackage[english]{babel}
\usepackage[applemac]{inputenc}

\usepackage{tikz}
\usetikzlibrary{decorations.pathmorphing}
\usepackage{stmaryrd}

\usepackage{color}

\begin{document}

\title{Diffusion 
 at the Surface of Topological Insulators}
\date{Version of \today}


\author{Pierre Adroguer}
\address{
Laboratoire de Physique, Ecole Normale Sup\'erieure de Lyon and CNRS UMR5672, France}

\author{David Carpentier}
\address{
Laboratoire de Physique, Ecole Normale Sup\'erieure de Lyon and CNRS UMR5672, France}

\author{J\'er\^ome Cayssol}
\address{
LOMA, University Bordeaux 1, F-33045 Talence, France}
\address{
Max-Planck-Institut f\"ur Physik komplexer Systeme, N\"othnitzer Str. 38, 01187 Dresden, Germany}
\address{
Department of Physics, University of California, Berkeley, California 94720, USA}

\author{Edmond Orignac}
\address{
Laboratoire de Physique, Ecole Normale Sup\'erieure de Lyon and CNRS UMR5672, France}

\begin{abstract}
We consider the transport properties of topological insulators surface states in
the presence of uncorrelated point-like disorder, both in the classical and quantum regimes. 
The
transport properties of those two-dimensional surface states depend strongly on the amplitude of the
hexagonal warping of their Fermi surface. It is shown that a perturbative analysis
of the warping fails to describe the transport in experimentally available topological insulators, such as 
 Bi$_2$Se$_3$ 
and Bi$_2$Te$_3$. Hence we
develop  a fully non-perturbative description of these effects. In particular, we find that the dependence of the
warping amplitude on the Fermi energy manifests itself in a strong dependence of
the diffusion constant on this Fermi energy, leading to several important
experimental consequences. 
Moreover, the combination of a strong warping with an in plane Zeeman effect
leads to an attenuation of conductance fluctuations in contrast to the situation of 
unwarped Dirac surface states.  
\end{abstract}
 
\maketitle
 

\section{Introduction}

Topological insulators (TIs) constitute a new state of matter with an insulating bulk and an odd number of metallic 
Dirac cones at their surface \cite{Hasan:2010,Qi:2011}. It was predicted 
\cite{Hzhang:2009} and confirmed by angle-resolved photoemission spectroscopy (ARPES) 
\cite{YXia:2009,Chen:2009,Hsieh:2009b} that the compounds Bi$_2$Se$_3$, Bi$_2$Te$_3$ and 
Sb$_2$Te$_3$ possess such a single surface state (SS).
Owing to this spin locking property, SSs of TIs are unique metallic 
systems for the study of fundamental magnetotransport properties and for the realization of 
future spintronics based devices.  
Unfortunately such transport experiments have faced a major difficulty so far. Indeed most TIs 
samples present enough residual bulk conductance to overwhelm the actual surface contribution
 \cite{Eto:2010,Analytis:2010,Butch:2010,Analytis:2010b}. The Shubnikov-de Haas oscillations 
reported in Bi$_2$Se$_3$ crystals originate exclusively from the bulk 3D bands indicating a 
low mobility for the surface Dirac fermions \cite{Eto:2010,Analytis:2010,Butch:2010}. 
Nevertheless recent progress has been achieved by using ultrathin films of Bi$_2$Se$_3$ 
 \cite{Checkelsky:2011,Steinberg:2010,Kim:2011} or Bi$_2$Te$_3$ \cite{Kong:2010} exfoliated 
on high-k dielectric insulator in order to improve the surface/bulk conductance ratio and to 
allow the gating of the surface state. From the gate voltage dependence of the total conductance, it was possible to 
separate bulk and surface contributions using a classical two-carrier model \cite{Checkelsky:2011,Steinberg:2010}.  
Most recently the surface conductance of strained HgTe samples has been 
reported in truly 3D slabs with thicknesses exceeding $100$ nm and in the absence of significant bulk conductance
\cite{Bouvier:2011}. 
At low temperatures and for small samples, the quantum correction to transport was recently 
probed experimentally : the dependence of conductance 
on weak magnetic fields displays the expected weak antilocalization while the 
Universal Conductance Fluctuations (UCF)
were observed 
in some of those experiments \cite{Kong:2010,Checkelsky:2011,Bouvier:2011}. 
Interestingly, the crossover between the symplectic and unitary universality classes 
upon breaking time-reversal symmetry 
has been observed in ultrathin samples of Bi$_2$Te$_3$ \cite{He:2011}.  

 Theoretically, describing the transport properties of topological insulator
 surface states amounts to consider the diffusion properties of two dimensional
 Dirac fermions. 
Indeed, at the lowest order in $k.p$ theory the SS Fermi 
surface is circular with a  spin winding in the plane.
At low energy, this two dimensional conductor shares a lot of similarities with graphene but 
with the following important differences: the momentum is locked to a real spin as opposed to a 
A-B sublattice pseudo-spin in graphene, and it has a single Dirac
cone as opposed to the four-fold degeneracy of the Dirac cone in graphene. For
this simpler Dirac metal,  the conductivity and the induced in-plane spin 
polarization were calculated as functions of the 2D carrier concentration \cite{Culcer:2010}. 
Far from the Dirac point,  surface 
spin-orbit may generically produce a significant hexagonal warping (HW) of the spin texture.
As a result the Fermi surface 
exhibits a snowflake or a nearly hexagonal Fermi surface depending on the carrier density, 
and the spin gets tilted 
out of the plane \cite{Fu:2009}. Those effects have been confirmed by ARPES and scanning tunneling microscopy 
(STM) experiments performed on Bi$_2$Te$_3$ crystals where HW is particularly strong 
\cite{Alpichshev:2010,Alpichshev:2011,Ando:2011,Xu:2011,Jung:2011}, and also in Bi$_2$Se$_3$ 
\cite{Kuroda:2010,Hirahara:2010}. Characterizing quantitatively the HW and therefore the 
amount of out-of plane 
spin polarization is a crucial issue in view of potential spintronics applications, and has been 
the subject of ab-initio studies 
\cite{Yazyev:2010} and ARPES experiments \cite{Xu:2011}. 
This hexagonal warping is specific to the Dirac metal at the surface of
topological insulators, and different from the trigonal warping encountered at
high energies in graphene. Its amplitude is stronger in the metallic regime of
high Fermi energy which is of interest in the present paper : a full description
of transport properties in this regime necessitates to include its presence in
the model. 
It has been shown \cite{Tkachov:2011,Wang:2011}  that 
hexagonal warping enhances perturbatively 
the classical Drude conductivity with respect to the unwarped model \cite{Culcer:2010}. 
 However, as we will show in this article a description of the effects of this
 hexagonal warping requires to go beyond the previous perturbative descriptions. In this paper, we develop a description of the diffusion of these 
surface states which is non-perturbative in the warping amplitude.   

We
 investigate theoretically the classical 2D charge diffusion within an hexagonally warped 
 surface states using both a standard diagrammatic formalism and a Boltzmann equation approach. 
 In this classical regime, we find that the hexagonal warping on one hand strongly reduces the density 
 of states and on the other hand increases the diffusion coefficient for the
 Dirac surface states. The combination of both effects is found to correspond to
an increase of the classical conductivity as a function of the warping amplitude.   
 Interestingly, as the warping amplitude increases with Fermi energy,
that dependence on the warping amplitude manifests itself in the  
behavior of the conductivity as a function of the Fermi energy. 
 In the second part of this article, we focus on the quantum correction to the
 conductivity, relevant in a phase coherent conductor. These quantum corrections
are known to depend on the symmetry class of the associated Anderson localization problem.  
Similarly to graphene with only intra-valley disorder, the single Dirac cone model is known to correspond to 
the symplectic / AII class. Within this symplectic class the Dirac model differs from graphene with
 long-range disorder or electrons with spin-orbit randomness by the presence of
a topological term in the associated field theory \cite{Schnyder:2008,Schnyder:2009}. 
 This topological term
explains the absence of Anderson localization for these topological insulators surface 
states \cite{Bardarson:2007,Nomura:2007,Ostrovsky:2010}, however 
it plays no role in the diffusive metallic regime. In this regime, the universal
properties of both weak localization and conductance fluctuations are just the
same as for any model on the two dimensional symplectic / AII class. We will
recall these results, and show that the dependence of the diffusion coefficient
on warping manifests itself in a dependence of the weak localization correction
and conductance fluctuations away from their universal values. In particular, this 
diffusion constant parameterizes the universal cross-over from the symplectic /
AII class to the unitary / A class when a magnetic field is applied
perpendicular to the surface states. The shape of the associated cross-over
functions now depends strongly on the warping amplitude, and thus the Fermi
energy.  Finally, we identify a property unique to the
topological insulator surface states : a reduction of the conductance
fluctuations by the application of a in-plane magnetic field.

The paper is organized as follows. 
 We present the model in section \ref{sec:model}. The classical Drude 
conductivity in presence of HW is derived using a Boltzmann approach 
(section \ref{sec:Boltzmann}) and a diagrammatic method (section \ref{sec:diagrams}). In section 
\ref{sec:quantum}, the quantum corrections to 
the average conductivity and the 
Universal Conductance Fluctuations are respectively derived. The results and conclusions relevants 
for experiments are 
summarized in the conclusive section \ref{sec:Results}.

\section{Model} 
\label{sec:model}

\subsection{Hamiltonian and Warping Potential}

We consider the Hamiltonian describing a single species of surface states  of a strong topological 
insulator
\begin{equation}
\label{eq:Dirac}
H = \hbar v_F ~\vec{\sigma} \cdot \vec{k} 
+ \frac{\lambda}{2}\sigma_{z}(k^3_+ +k^3_-)
+ V(\vec{r}), 
\end{equation} 
where 
$\vec{\sigma}=(\sigma_x,\sigma_y)$ is the two-dimensional vector of Pauli matrices describing the physical 
in-plane spin
of the electron, and $\vec{k}=(k_x,k_y)$ is
its in-plane momentum operator which satisfy the 
usual commutation relation with the position operator 
$[r_\alpha,k_\beta]=i\delta_{\alpha\beta}$ ($\alpha,\beta=x,y,z$).
Note that in Eq.(\ref{eq:Dirac}), we used for sake of simplicity the dispersion relation 
$\vec{\sigma} \cdot \vec{k} $ for the Dirac part instead of the usual $\vec{\sigma} \times \vec{k} $
 hamiltonian : switching from one to the other amounts to perform an in plane $\pi/2$ rotation around $z$ in 
 the spin  space and has no consequence on the following discussion as it does not affect the
 hexagonal warping potential. 
Concerning the description of the disorder potential $V(\vec{r})$, 
we will follow the standard description  by considering a Gaussian random potential $V(\vec{r})$, 
characterized by a zero average value 
$\langle V(\vec{r}) \rangle = 0$ and variance $\langle V(\vec{r})V(\vec{r'}) \rangle = \gamma \delta(\vec{r}-\vec{r'})$,
where $\langle \dots \rangle$ represents the average over different disorder realizations. 

\begin{figure} [!h]
\centering 
\includegraphics[width=0.4 \textwidth]{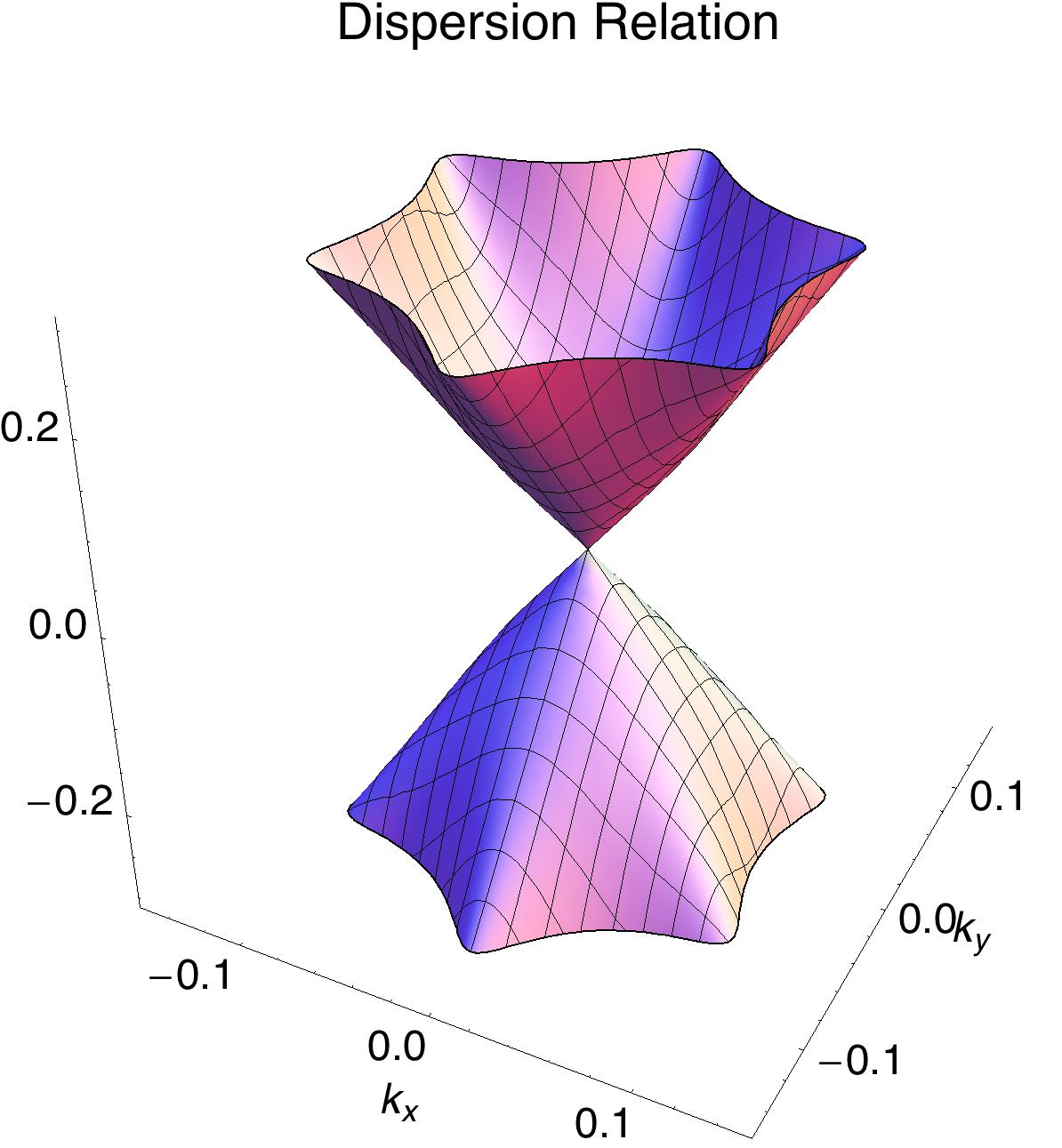} 
\hspace{.5cm}
\includegraphics[width=0.4 \textwidth]{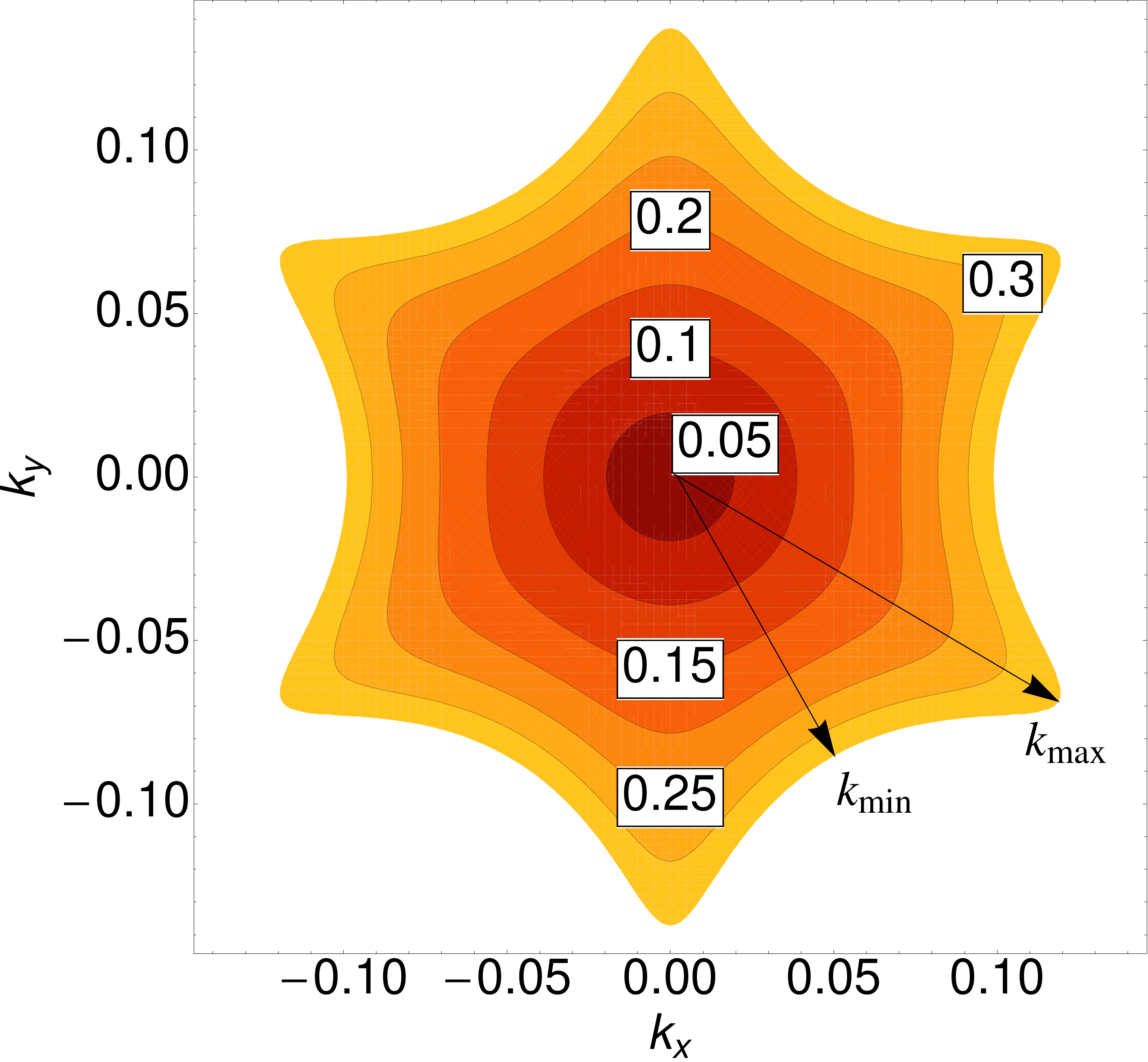} 
\caption{Fermi surface in presence of hexagonal warping for various Fermi energies with parameters relevant for 
Bi$_{2}$Te$_{3}$ : 
$\lambda = 250\ \textrm{eV}.\mathring{A}^3$ and $v_F=2.55 \ \textrm{eV}.\mathring{A}$ and Fermi energies 
from $E_{F}=0.05$ eV to $0.3$ eV, corresponding to a warping parameter $b$ ranging from $0$ to $b\simeq 0.7$.}
\label{fig:energy}
\end{figure}

The hexagonal warping term 
$H_{\textrm{w}}= \frac{\lambda}{2}\sigma_{z}(k^3_+ +k^3_-)$, with $k_{\pm} = k_x \pm i k_y$, breaks the full 
$U(1)$ rotation symmetry of the Dirac cone down to a trigonal discrete $C_3$ symmetry \cite{Fu:2009,Liu:2010}. 
Time reversal symmetry leads to an additional hexagonal $C_6$ symmetry of the Fermi surface, hence the name 
hexagonal warping (Fig.\ref{fig:energy}). The dispersion relation in the absence of disorder with
$\vec{k}=(k_x,k_y)=k(\cos \theta,\sin \theta)$ :
\begin{equation}
\label{eq:Dispersion}
\epsilon^{2}(\vec{k}) =  \hbar^2 v_F^2 k^2 + \lambda^2 k^6 \cos^2 (3 \theta) , 
\end{equation} 
is obtained by squaring the Hamiltonian Eq.(\ref{eq:Dirac}) with $V(\vec{r})=0$, and leads to the snowflake shape of 
equal energy surfaces at high energies : see Fig.\ref{fig:energy}. 
Defining $E_{F}=\hbar v_{F} k_{F}$ and  
$k=k_{F} \tilde{k}(\theta)$, the shape of the Fermi surface at energy $E_F$  is defined by the equation 
\begin{equation}
\label{eq:Dispersion2}
1 =  \tilde{k}^2(\theta) + 4 b^{2}  \tilde{k}^6(\theta) \cos^2 (3 \theta). 
\end{equation} 
Hence the warping  of this Fermi surface is naturally characterized by the dimensionless parameter 
\begin{equation}
b= \frac{\lambda E_F^2}{ 2\hbar^3 v_F^3},
\end{equation}
 which will play a crucial role in the remaining of this paper. 
This parameter is related to geometrical  warping deformation factor $w$ introduced in ref.~\cite{Xu:2011}, and 
defined from the maximum and minimum momenta $k_{max},k_{min}$ of a constant energy contour (see 
Fig.\ref{fig:energy}) as  
\begin{equation}
\tilde{w}  = \frac{w}{w_{max}} = \frac{k_{max}-k_{min}}{k_{max}+k_{min}}
\quad
\textrm{ with }
\quad 
w_{max} = \frac{2+\sqrt{3}}{2-\sqrt{3}}  \simeq 13.9
\end{equation} 
With the notations of Eq.~(\ref{eq:Dispersion2}), we have $k_{min}/k_{max} = \tilde{k}(\theta=0)$, leading to the 
expression  
\begin{equation}
b(w) = \frac{\sqrt{\tilde{w}} (1+\tilde{w})^{2}}{(1-\tilde{w})^{3}}= 
\sqrt{w w_{max}} \frac{(w_{max}+w)^{2}}{(w_{max} - w)^{3}} . 
\end{equation} 
Two values appear remarkable : 
 $w=0, b=0$  correspond to a circular Fermi Surface ;  
$w=1,b=2/(3\sqrt{3})\simeq 0.38$ to a hexagonal Fermi surface; while $w>1$ indicates a snowflake shape. 
  
  Using the experimental values for the Bi$_2$Se$_3$ compound
 $\lambda = 128\ \textrm{eV}.\mathring{A}^3$ , $v_F=3.55 \ \textrm{eV}.\mathring{A}$ from \cite{Kuroda:2010}
we obtain relatively small values of warping  $0.04 < b < 0.09$ for energies   $0.05 eV < E < 0.15 eV$, 
and similarly $0.0 < b < 0.04$ 
with the values $\lambda = 95\ \textrm{eV}.\mathring{A}^3$ , $v_F=3.0 \ \textrm{eV}.\mathring{A}$  
and   $ 0.00 \mathrm{eV} < E < 0.15 \mathrm{eV}$ from \cite{Hirahara:2010}.   
On the other hand, the experimental values for the   
  Bi$_{2}$Te$_{3}$ compound 
$\lambda = 250\ \textrm{eV}.\mathring{A}^3$ and $v_F=2.55 \ \textrm{eV}.\mathring{A}$
\cite{Chen:2009,Alpichshev:2010}
lead to a warping factor ranging from $b=0.13$ for $E=0.13$ eV to $b= 0.66$ for $E=0.295$ eV. As 
we will show below, these large values of warping amplitude are indeed beyond the reach of a perturbative approach, 
and require
a non-perturbative description of diffusion in the presence of warping.

\section{Classical Drude conductivity: Boltzmann equation}
\label{sec:Boltzmann}

The momentum relaxation rate in presence of warping can be obtained
from a simple application of  Fermi's Golden Rule:
\begin{equation}
\frac{1}{\tau_e} = \int \! \frac{d\vec{k'}}{(2 \pi)^2} 2 \pi \vert \langle \vec{k'} \vert \vec{k} \rangle \vert^2 \gamma 
\delta (\epsilon(\vec{k'})-E_F),
\label{eq:golden}
\end{equation} 
where $\langle \vec{k'} \vert \vec{k} \rangle$ is the overlap between the eigenstates labelled by $\vec{k}$ and 
$\vec{k'}$ at the Fermi surface. The disorder strength is characterized by $\gamma = n_i V^2$, and the dispersion 
$\epsilon(\vec{k})$ is given by Eq. (\ref{eq:Dispersion}). 

\begin{figure} [!h]
\centering 
\includegraphics[width=0.3 \textwidth]{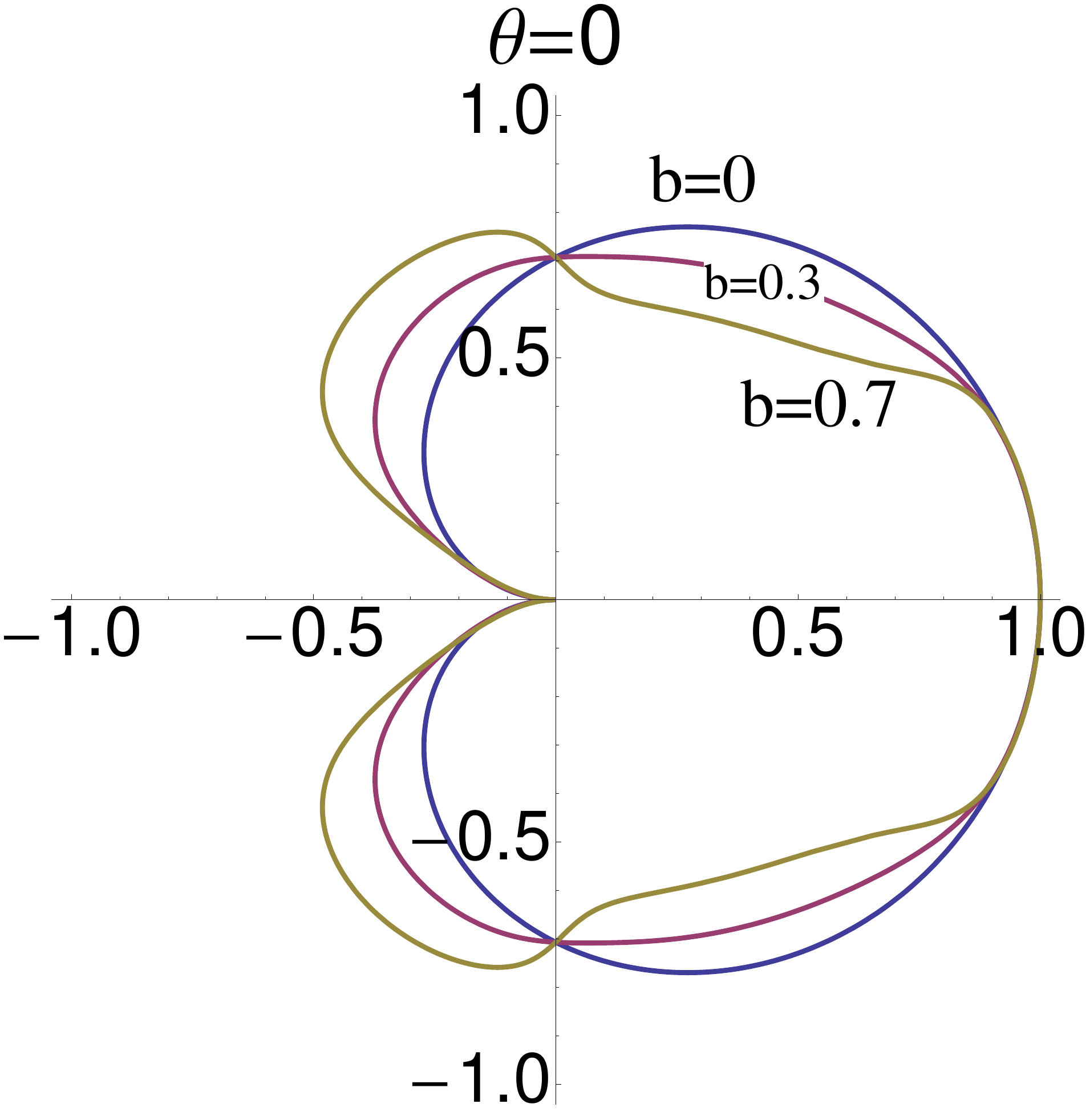} 
\includegraphics[width=0.3 \textwidth]{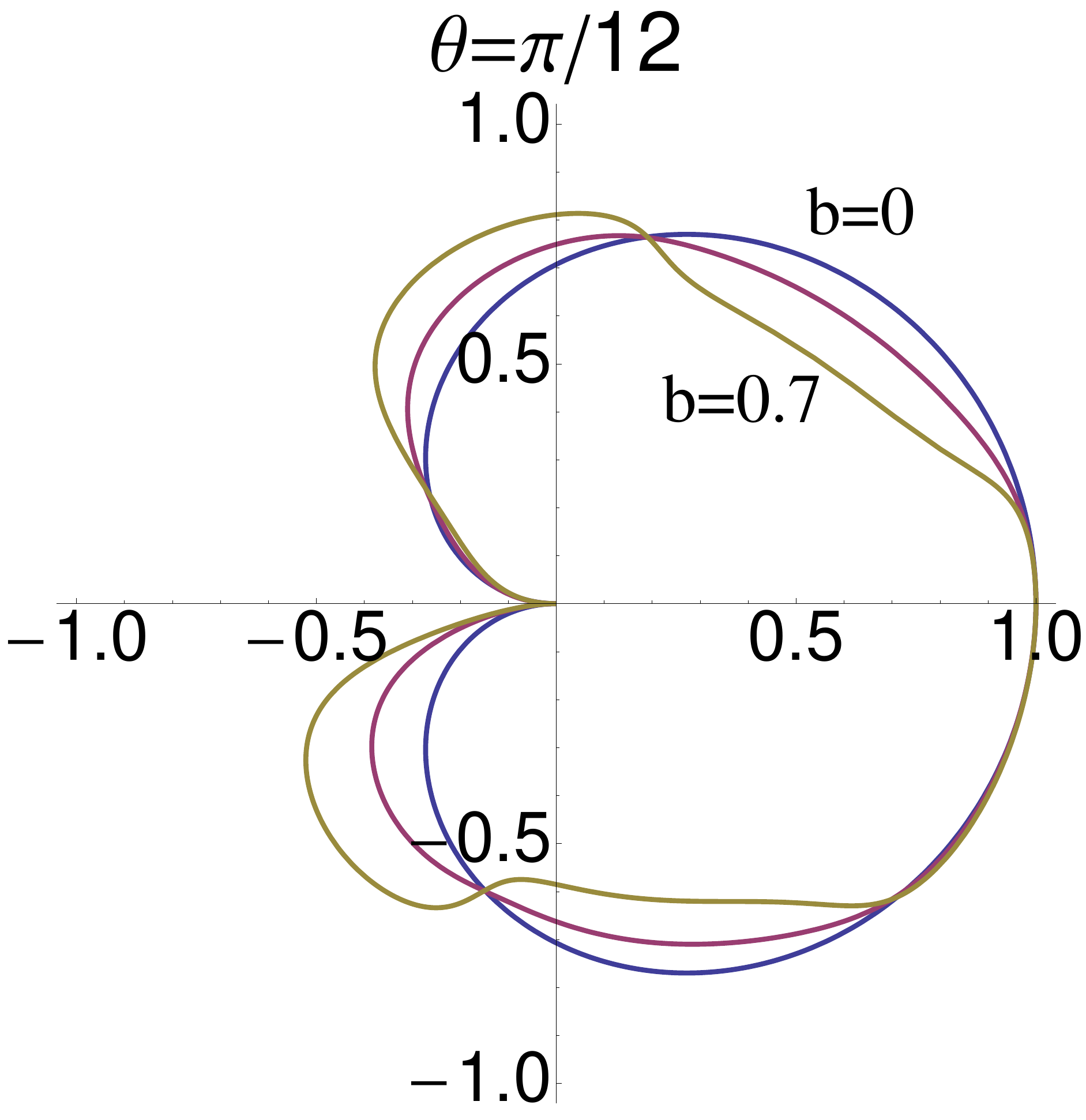} 
\includegraphics[width=0.3 \textwidth]{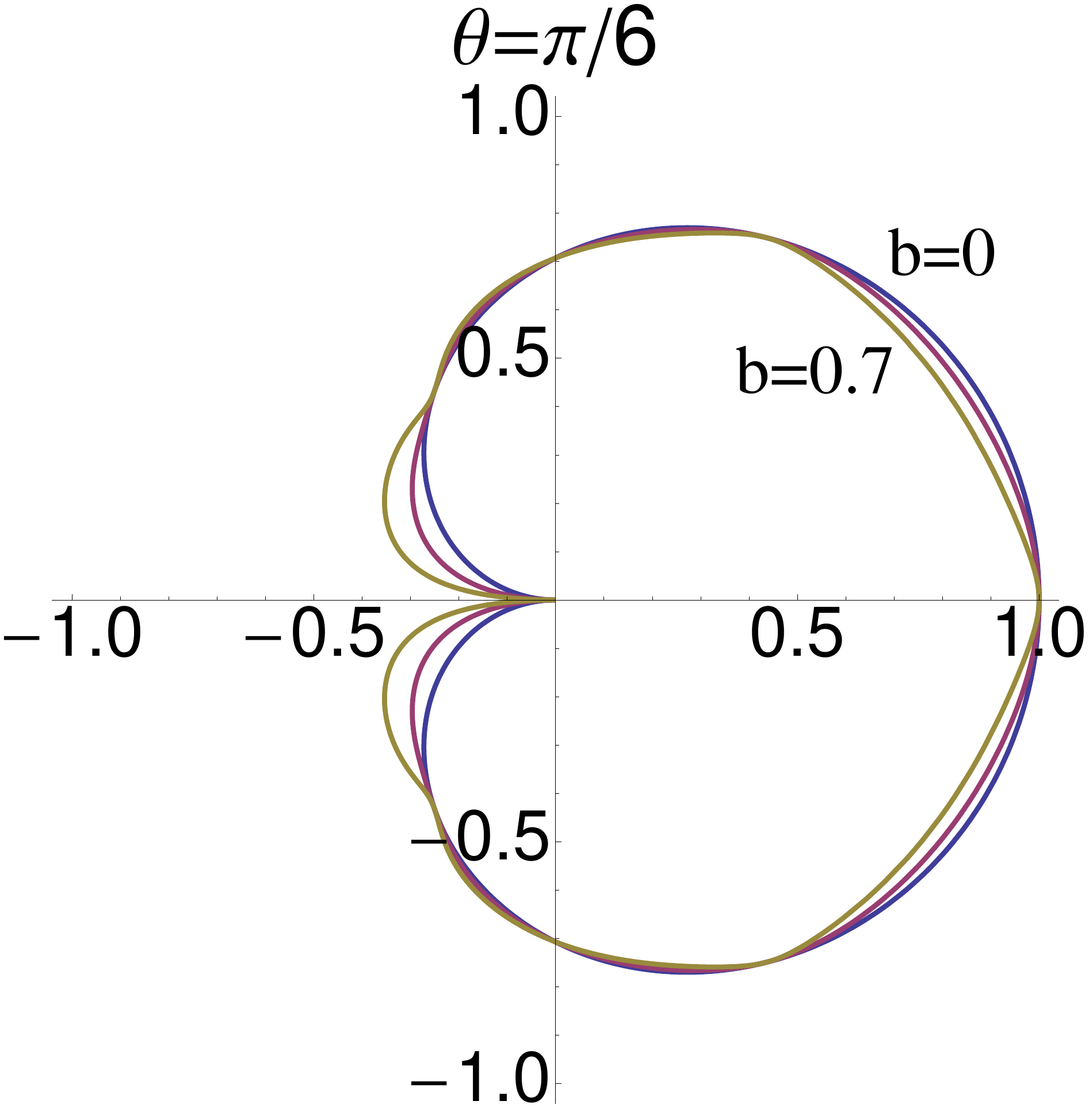} 
\caption{Scattering amplitude $|\langle \vec{k}|V|\vec{k'}\rangle|^{2}$ plotted as a function of the relative angle 
$\theta'-\theta$ between the incident state $\vec{k}$ of polar angle $\theta$ and the scattered state $\vec{k}'$ of 
polar angle $\theta'$. Three different incident directions are plotted to demonstrate the dependence on the polar 
angle of the incident state :  
$\theta=0,\pi/12,\pi/6$, while the results are invariant by $\pi/3$ rotation of the incident state.
The results are shown for increasing energies warping factor $b=0-0.7$ corresponding to increasing Fermi energies
 in a given material. 
}
\label{fig:anisotropy}
\end{figure}

At equilibrium, the occupation numbers are given by the Fermi distribution $n_F(\epsilon(\vec{k}))$. In presence of 
a finite external electric field $\vec{E}$, the occupation number function $f(\vec{k})$ has to be determined by solving
 the Boltzmann equation:
\begin{equation}
\label{eq:Boltzmann}
-e \vec{E} . \frac{\partial f(\vec{k})}{\partial \vec{k}} = \int \! \frac{d \vec{k'}}{(2 \pi)^2} 2 \pi n_i \vert 
\langle \vec{k'} \vert V \vert \vec{k} \rangle \vert^2 \delta(\epsilon(\vec{k'})-\epsilon(\vec{k})) 
\left[ f(\vec{k'}) - f(\vec{k}) \right] .
\end{equation}
This equation is nontrivial to solve due to both the warped shape of the Fermi surface and the anisotropy of 
the scattering. Indeed, even if the scalar disorder is isotropic and spin diagonal, the spin-locking property makes the 
scattering anisotropic (even for a circular unwarped Fermi surface). This anisotropy is further increased by the 
hexagonal warping of the Fermi surface. 
This is apparent when considering the scattering amplitude which appears in Eq.(\ref{eq:Boltzmann}) : 
$f(\theta, \theta'-\theta)=|\langle \vec{k}|V|\vec{k'}\rangle|$ between eigenstates of the disorder-free Hamiltonian 
$H_{0}+H_{\textrm{w}}$ in Eq.(\ref{eq:Dirac}), as a function of the polar angle $\theta$ on the incident state 
$|\vec{k}\rangle$ and polar angle $\theta'$ of the outgoing  state $|\vec{k}'\rangle$. Due to the presence of 
hexagonal warping, this scattering amplitude is highly spin-dependent and anisotropic along the Fermi surface, 
especially at large Fermi energy where the warping factor $b$ is large. 
Moreover this amplitude $f(\theta, \theta'-\theta)$ becomes dependent on the incident state direction $\theta$,
 and not only on the relative angle, $(\theta'-\theta)$ as opposed to the case of Dirac fermions without warping. 
 These properties are represented in 
Fig. \ref{fig:anisotropy}.

\begin{figure} [!h]
\centering 
\includegraphics[width=8cm]{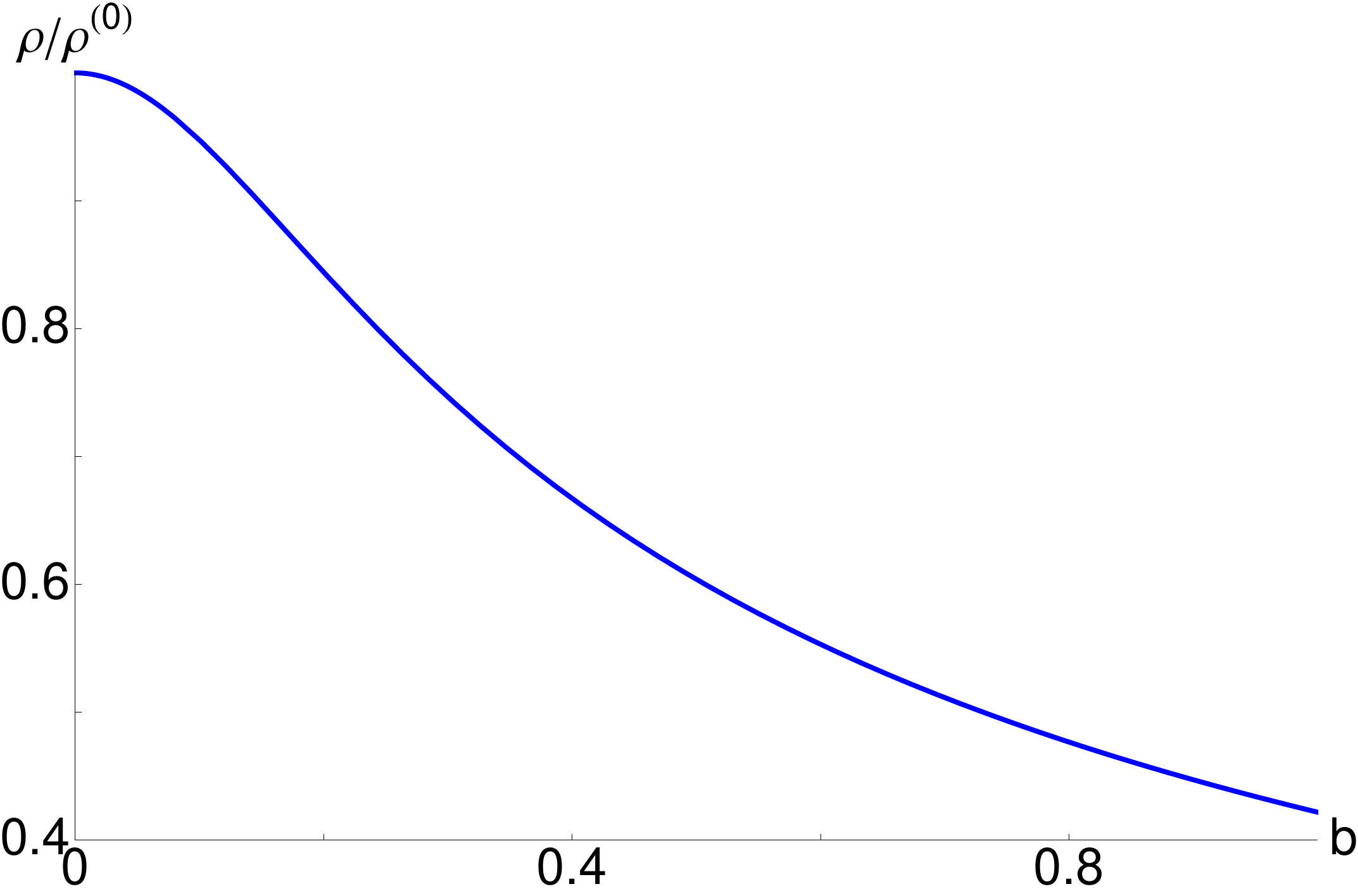} 
\caption{
Renormalization of the density of states $\rho(E_{F})$ by the warping of the surface states, non perturbatively in the 
warping amplitude $b$.
 The resulting correction is represented $\rho(E_{F})/\rho^{(0)}$ where $\rho^{(0)}=\rho(\lambda=0)$ is 
 the  density for Dirac Fermions without warping. }
\label{fig:densitystates-b}
\end{figure}

To solve the equation (\ref{eq:Boltzmann}), we use the  linearized ansatz:
\begin{equation}
f(\vec{k}) = n_F\left(\epsilon(\vec{k})\right) + \frac{\partial n_F}{\partial \epsilon} \bar{f} (\theta),
\end{equation}
where the angular function $\bar{f} (\theta)$ is linear in electric field. We find:
\begin{equation}
-e \vec{E}. \frac{\partial \epsilon}{\partial \vec{k}} = 2 \pi \gamma \int \! \frac{d \theta'}{(2 \pi)^2} \vert  \langle 
\vec{k'} \vert \vec{k} \rangle \vert^2 k_{F\theta'} \left( \frac{\partial \epsilon}{\partial k}  \right)_{\theta'}^{-1}  
\left[ \bar{f}(\theta') - \bar{f}(\theta)\right],
\end{equation}
where the Fermi wavevector $k_{F\theta'}$ and the derivative $(\partial \epsilon /\partial k)_{\theta'}$ are both 
evaluated at the point of the Fermi surface labelled by the angle $\theta'$. The solution of this integral equation at 
fourth order in $b$ is:
\begin{eqnarray}
\bar{f}(\theta) = & e v_F \tau_e E_x 
\biggl[2 \cos \theta + b^2 \left(18 \cos \theta + 5 \cos (5 \theta) - \cos (7 \theta)\right) 
\\
\nonumber 
&
 -\frac{b^4}{4} \left(522 \cos \theta + 20 \cos (5 \theta) + 44 \cos (7 \theta) + 49 \cos (11 \theta) - 11 \cos (13 \theta)\right) 
 \biggr].
\end{eqnarray}
The current is then calculated as:
\begin{equation}
j_{x} = \int \! \frac{d \vec{k}}{(2 \pi)^2}  e  \frac{\partial \epsilon}{\partial k_{x}}  \bar{f}(\theta)
\delta (\epsilon(\vec{k}) - E_F) =\sigma E_{x} ,
\end{equation}
leading to the conductivity:
\begin{equation}\label{eq:perturb-sigma}
\sigma =  \sigma^{(0)}  \left( 1 + 8b^2 - 58 b^4 + o(b^4) \right).
\end{equation} 
The conductivity and density of states (at Fermi level) in the absence of warping ($b=0$) are given respectively by 
$\sigma^{(0)} = e^2 \rho^{(0)} v_F^2 \tau^{(0)}_e $ and $\rho^{(0)} = \frac{E_F}{2 \pi \hbar^{2} v_F^{2}}$.
Similarly, one can also derive the perturbative (and non-perturbative) expression for the density of state $\rho$ : this is done in 
 \ref{app:density} and the result is represented on Fig.~\ref{fig:densitystates-b}. However, to go beyond the 
above perturbative expansion for the conductivity, a diagrammatic approach turns out to be more convenient. Hence we proceed below by developing such an approach, comparing its results with the perturbative expansion of Eq. (\ref{eq:perturb-sigma}). 

\section{Classical Drude conductivity : diagrammatic approach}
\label{sec:diagrams}

In this part, we investigate within the standard diagrammatic framework the  diffusive transport of Dirac surface state with an 
arbitrary large warping deformation (see Eq.~(\ref{eq:Dirac})). We start by calculating the single particle Green function 
averaged over disorder whose imaginary part yields the density of
states. Then  we evaluate the classical Drude 
conductivity from the bubble diagram containing two dressed current operators linked by two disorder averaged 
Green functions. We discuss the dependence of the classical conductance as a function of the 2D carrier density and 
the HW coupling strength.

\subsection{Averaged Green's function and elastic scattering time.}

The averaged Green's function is obtained by calculating the self-energy correction and averaging over disorder  \cite{McCann:2006}. 
The corresponding disorder averaged propagator reads: 
\begin{equation}
\langle G^{R/A}(\vec{k}) \rangle = \frac{(E\pm i \hbar/ 2 \tau_e) \mathbb{1} + \hbar v_F \vec{k} \ . \ \vec{\sigma} 
+ \lambda k^3 \cos(3 \theta) \sigma^z}{(E\pm i \hbar/ 2 \tau_e)^2 - \hbar^2 v_F^2 k^2 - \lambda^2 k^6 
\cos^2( 3\theta)}
\label{eq:Gwarp}
\end{equation} 
where the elastic scattering time for the warped (resp. unwarped) conical Dirac fermion 
$\tau_e$ (resp. $\tau_e^{(0)})$ is defined by :
\begin{equation}
\rho \tau_e  = \rho^{(0)} \tau^{(0)}_e  = \frac{\hbar}{\pi \gamma}. 
\label{eq:tauwarp}
\end{equation} 
We assume that the phase coherence time $\tau_\phi
\gg \tau_e$ in order to be in the regime of coherent transport. 
By using the parameterization $\tilde{k}(\theta)$ of the Fermi surface shape introduced in Eq.~(\ref{eq:Dispersion2}), 
we obtain the non perturbative expression for the the density of states  : 
\begin{equation}
  \frac{\rho}{\rho^{(0)}} = \alpha(b) = \int_{0}^{2\pi} \! \frac{d \theta}{2 \pi} \frac{1}{1 + 12~ b^2 \tilde{k}^4(\theta) 
  \cos^2(3 \theta) } = \frac{\tau_e^{(0)}}{\tau_e}
\label{eq:alpha}
\end{equation}
 We find an increase of the scattering time $\tau_e $ by the warping. 
Hence in the presence of hexagonal warping, the surface states of topological insulators are relaxing carrier 
momentum less effectively than without warping. This is a manifestation of the increased anisotropy of the 
scattering amplitude by the hexagonal warping correction, illustrated 
 in Fig.\ref{fig:anisotropy}. Indeed in the presence of hexagonal warping, the neighboring states of an incident Dirac 
 fermions have smaller overlaps than states that are $\pm 2\pi/3$ apart, which effectively induces a very anisotropic 
 scattering. 
Nevertheless from this result, one can not yet conclude that the Drude conductance is enhanced as the density of 
states at the Fermi energy is also renormalized by the warping.  
Indeed the Drude conductance is determined by the diffusion constant $D$ 
and the transport relaxation time rather than the elastic scattering time $\tau_e $ governing momentum relaxation. 
To access this classical conductivity, we follow the standard diagrammatic procedure in the following section. 

\subsection{Diagrams for the classical conductivity} 

For a given impurity configuration, the  conductivity is given by the Kubo formula :
\begin{equation}
\sigma = \sigma_{x x} = \frac{\hbar}{2\pi L^2} \Re \mathrm{Tr} \left[ j_{x} G^{R}j_{x}  G^{A} \right],
\end{equation} 
where we have neglected subdominant contributions  involving products $G^{R} G^{R}$ and $ G^{A} G^{A}$ 
\cite{Akkermans:2007}. 
In this expression and below, $\mathrm{Tr}$ denotes a trace over the electron's Hilbert space (momentum and spin 
quantum numbers). 
The current operator $j_{\alpha}$ is obtained from the Hamiltonian Eq. (\ref{eq:Dirac}) by inserting the vector potential 
{\it via} the minimal coupling substitution $\vec{k} \to \vec{k}  - (e/\hbar) \vec{A}$  
and using the definition  $j_{\alpha} =  \frac{\delta H}{\delta A_{\alpha}}$:
\begin{equation}
j_{x}  = (-e) \left(  v_{F}~ \sigma_x + \frac{3 \lambda}{ \hbar} \sigma_z (k_x^2-k_y^2) \right).
\end{equation} 
Note that this current contains the usual $-e v_{F} \sigma_{x}$ term for linearly dispersing Dirac system, and an 
additional  term  quadratic in momentum originating from the HW. 
  
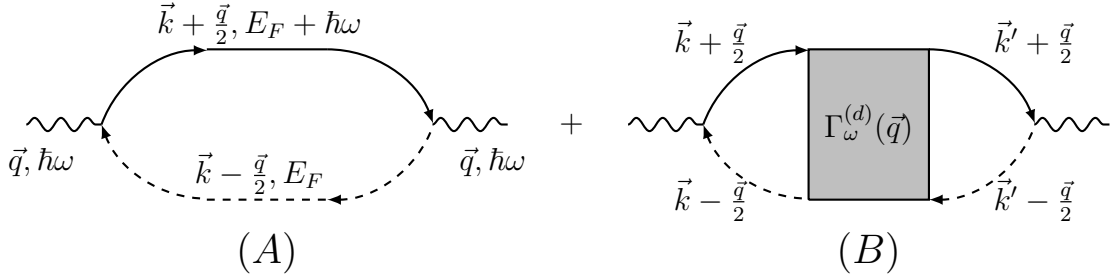
\begin{figure} [!th]
\centering 
\begin{tikzpicture}
[decoration=snake,line around/.style={decoration={pre length=#1,post length=#1}},scale=1]
\begin{scope}
\draw[thick] (4.9,1) -- (6.5,1) ;
\draw[thick,dashed] (4.9,-1) -- (6.5,-1) ;
\end{scope}
\draw[-,decorate,  thick] (2.5,0) -- (3.5,0);
\draw[-,decorate,  thick] (7.9,0) -- (8.9,0);
\begin{scope}
\draw[->, >=latex, thick] (3.5,0) arc (160:90:1.5);
\draw[->, >=latex, thick] (6.5,1) arc (90:20:1.5);
\draw[<-, >=latex,dashed,  thick] (6.5,-1) arc (270:340:1.5);
\draw[<-, >=latex , dashed,  thick] (3.5,0) arc (200:270:1.5);
\end{scope}
\node at (2.7,-0.5) {$\vec{q},\hbar \omega$};
\node at (8.7,-0.5) {$\vec{q},\hbar \omega$};
\node at (5.6,1.35) {$\vec{k}+\frac{\vec{q}}{2},E_{F}+\hbar \omega$};
\node at (5.6,-0.65) {$\vec{k}-\frac{\vec{q}}{2},E_{F}$};
\node[scale=1.3] at (5.7,-1.7) {$(A)$};
\node at (9.75,0.) {$+$};
\begin{scope}
\draw[thick,fill=black!50,fill opacity=0.5] (12.9,-1) node{} 
-- (12.9,1) node{} 
-- (14.5,1) node{}
-- (14.5,-1) node{} 
--(12.9,-1) node{};
\node[scale=1] at (13.7,0) {$\Gamma^{(d)}_{\omega}(\vec{q})$};
\node[scale=1.3] at (13.7,-1.7) {$(B)$};
\end{scope}
\draw[-,decorate,  thick] (10.5,0) -- (11.5,0);
\draw[-,decorate,  thick] (15.9,0) -- (16.9,0);
\begin{scope}
\draw[->, >=latex, thick] (11.5,0) arc (160:90:1.5);
\draw[->, >=latex, thick] (14.5,1) arc (90:20:1.5);
\draw[<-, >=latex,dashed,  thick] (14.5,-1) arc (270:340:1.5);
\draw[<-, >=latex , dashed,  thick] (11.5,0) arc (200:270:1.5);
\end{scope}
\node at (11.6,1.1) {$\vec{k}+\frac{\vec{q}}{2}$};
\node at (11.6,-1.) {$\vec{k}-\frac{\vec{q}}{2}$};
\node at (15.9,1.1) {$\vec{k}'+\frac{\vec{q}}{2}$};
\node at (15.9,-1.) {$\vec{k}'-\frac{\vec{q}}{2}$};
\end{tikzpicture}
\caption{Diagrammatic representation of the two contributions $\sigma_{A}$ and $\sigma_{B}$ 
to the classical conductance}
\label{fig:diagramsClassical}
\end{figure}

After averaging over all impurity configurations, the classical mean conductivity 
is obtained from the bubble diagram of Fig.~\ref{fig:diagramsClassical} where the propagating lines represent the retarded 
and advanced disorder averaged Green functions Eq. (\ref{eq:Gwarp}). This classical conductivity is the sum of two 
terms : $\sigma_{A}$ and $\sigma_{B}$.  
%
The contribution of diagram A in Fig. \ref{fig:diagramsClassical} is defined at $\vec{q}=0$ and $\omega=0$ by 
\begin{eqnarray}
\sigma_{A} = \frac{\hbar}{2 \pi} 
\int \frac{d\vec{k}}{(2 \pi)^2} 
~\mathrm{tr}
& \left[  
j_x (\vec{k}) 
\langle G^{R}(\vec{k},E_{F}) \rangle 
 j_x (\vec{k})
\langle G^{A}(\vec{k},E_{F}) \rangle \right] ,
\end{eqnarray}
where $\mathrm{tr}$ denotes a trace only over the spin quantum numbers. 
 Performing explicitly the trace and the integration over the momentum $k=|\vec{k}|$, we obtain the following 
 expression: 
\begin{equation}
\sigma_A =   \frac{\hbar e^2 v_F^2}{ 2 \pi \gamma} \frac{\alpha(b) +5 \beta(b) + \delta(b)}{\alpha(b)} , 
\label{eq:sigma_A}
\end{equation} 
which is non perturbative in $b$. The function $\alpha(b)$ has been defined in Eq.(\ref{eq:alpha}) and we have introduced :
\begin{eqnarray}
\alpha(b) &=& \int_{0}^{2\pi} \! \frac{d \theta}{2 \pi} \frac{1}{1 + 12~ b^2 \tilde{k}^4(\theta) \cos^2(3 \theta) }\\
\beta(b) &=& \int_{0}^{2\pi} \frac{d \theta}{2 \pi} \frac{4b^2\cos^2(3 \theta) \tilde{k}^6(\theta)}{1 + 12 b^2 
\cos^2(3 \theta) \tilde{k}^4(\theta)},  \\
\delta(b) &=& \int_{0}^{2\pi} \frac{d \theta}{2 \pi} \frac{36b^2(\tilde{k}^4(\theta) -\tilde{k}^6(\theta))}{1 + 12 b^2 
\cos^2(3 \theta) \tilde{k}^4(\theta)} , 
\end{eqnarray}
where $k=k_{F} \tilde{k}(\theta)$ with $ \tilde{k}(\theta)$ introduced in Eq.~(\ref{eq:Dispersion2}). 

The contribution of the diagram B in Fig. \ref{fig:diagramsClassical} accounts for the contribution of the so-called 
Diffuson 
\cite{Akkermans:2007}. Retaining only the dominant contribution of the integral, we can write it explicitly as 
\begin{equation}
\sigma_{B}(\vec{q},\omega) = 
\frac{\hbar}{2 \pi} \mathrm{tr} \left[ J_x \Gamma^{(d)}(\vec{q},\omega) J_x \right],
\label{eq:sigmaB}
\end{equation}
 where the vertex operator $J$ is defined by 
\begin{equation}
J_x =  \int \frac{d\vec{k}}{(2 \pi)^2} 
\langle G^{A}(\vec{k},E) \rangle \ j_x(\vec{k}) \ \langle G^{R}(\vec{k},E) \rangle   = J ~\sigma_{x} .
\label{eq:J}
\end{equation}
A contraction over the spin indices is assumed, resulting in the proportionality to $\sigma_{x}$. 

\begin{figure}
\begin{center}
\begin{tikzpicture}
[decoration=snake,
line around/.style={decoration={pre length=#1,post length=#1}},scale=0.8]
\begin{scope}
\draw[thick,fill=black!50,fill opacity=0.5] (0,0) node{} -- (0,1) node{} -- (2,1) node{} -- (2,0) node{} --(0.,0) node{};
\node[scale=1] at (1,0.5) {$\Gamma^{(d)}$};
\end{scope}
\node[scale=1] at (2.5,0.5) {$=$};
\draw[-,  thick] (3,0) -- (3.2,0);
\draw[-,  thick] (3,1) -- (3.2,1);
\draw[-,  dotted] (3.1,0) -- (3.1,1);
\node[scale=1] at (3.1,0.5) {$\times$};
\node[scale=1] at (3.7,0.5) {$+$};
\draw[-,  thick, dashed] (4.4,0) -- (5.1,0);
\draw[-,  thick] (4.4,1) -- (5.1,1);
\draw[-,  dotted] (4.5,0) -- (4.5,1);
\draw[-,  dotted] (5.,0) -- (5.0,1);
\node[scale=1] at (6.3,0.5) {$+$};
\node[scale=1] at (4.5,0.5) {$\times$};
\node[scale=1] at (5.0,0.5) {$\times$};
\draw[-,  thick, dashed] (7.4,0) -- (8.6,0);
\draw[-,  thick] (7.4,1) -- (8.6,1);
\draw[-,  dotted] (7.5,0) -- (7.5,1);
\draw[-,  dotted] (8.0,0) -- (8.0,1);
\draw[-,  dotted] (8.5,0) -- (8.5,1);
\node[scale=1] at (10.1,0.5) {$+$};
\node[scale=1] at (7.5,0.5) {$\times$};
\node[scale=1] at (8.0,0.5) {$\times$};
\node[scale=1] at (8.5,0.5) {$\times$};
\draw[-,  thick, dashed] (11.4,0) -- (13.1,0);
\draw[-,  thick] (11.4,1) -- (13.1,1);
\draw[-,  dotted] (11.5,0) -- (11.5,1);
\draw[-,  dotted] (12.0,0) -- (12.0,1);
\draw[-,  dotted] (12.5,0) -- (12.5,1);
\draw[-,  dotted] (13.0,0) -- (13.0,1);
\node[scale=1] at (14.5,0.5) {$+ \cdots$};
\node[scale=1] at (11.5,0.5) {$\times$};
\node[scale=1] at (12.0,0.5) {$\times$};
\node[scale=1] at (12.5,0.5) {$\times$};
\node[scale=1] at (13.0,0.5) {$\times$};
\end{tikzpicture}
\begin{tikzpicture}
[decoration=snake,
line around/.style={decoration={pre length=#1,post length=#1}},scale=0.8]
\begin{scope}
\draw[thick,fill=black!50,fill opacity=0.5] (0,0) node{} -- (0,1) node{} -- (2,1) node{} -- (2,0) node{} -- (0.,0) node{};
\node[scale=1] at (1,0.5) {$\Gamma^{(c)}$};
\end{scope}
\node[scale=1] at (2.5,0.5) {$=$};
\draw[-,  thick, dashed] (4.4,0) -- (5.1,0);
\draw[-,  thick] (4.4,1) -- (5.1,1);
\draw[-,  dotted] (4.5,0) -- (5.0,1);
\draw[-,  dotted] (5.,0) -- (4.5,1);
\node[scale=1] at (4.75,0.5) {$\times$};
\node[scale=1] at (6.3,0.5) {$+$};
\draw[-,  thick, dashed] (7.4,0) -- (8.6,0);
\draw[-,  thick] (7.4,1) -- (8.6,1);
\draw[-,  dotted] (7.5,0) -- (8.5,1);
\draw[-,  dotted] (8.0,0) -- (8.0,1);
\draw[-,  dotted] (8.5,0) -- (7.5,1);
\node[scale=1] at (8.,0.5) {$\times$};
\node[scale=1] at (10.1,0.5) {$+$};
\draw[-,  thick, dashed] (11.4,0) -- (13.1,0);
\draw[-,  thick] (11.4,1) -- (13.1,1);
\draw[-,  dotted] (11.5,0) -- (13.0,1);
\draw[-,  dotted] (12.0,0) -- (12.5,1);
\draw[-,  dotted] (12.5,0) -- (12.0,1);
\draw[-,  dotted] (13.0,0) -- (11.5,1);
\node[scale=1] at (12.25,0.5) {$\times$};
\node[scale=1] at (14.5,0.5) {$+ \cdots$};
\end{tikzpicture}
\end{center}
\caption{\label{fig:DiffusonPropag} Schematic representation of the Diffuson and Cooperon structure factors. 
The dotted lines correspond to disorder correlators, while the plain and dashed lines represent retarded and 
advanced Green's functions. 
 }
\end{figure}
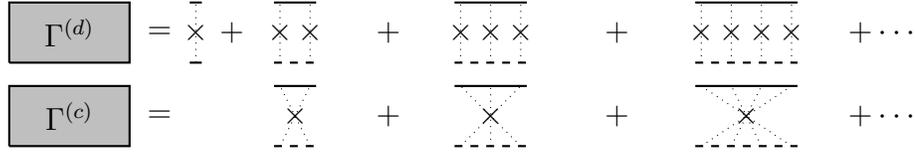
\begin{figure} [!h]
\centering 
\includegraphics[width=10cm]{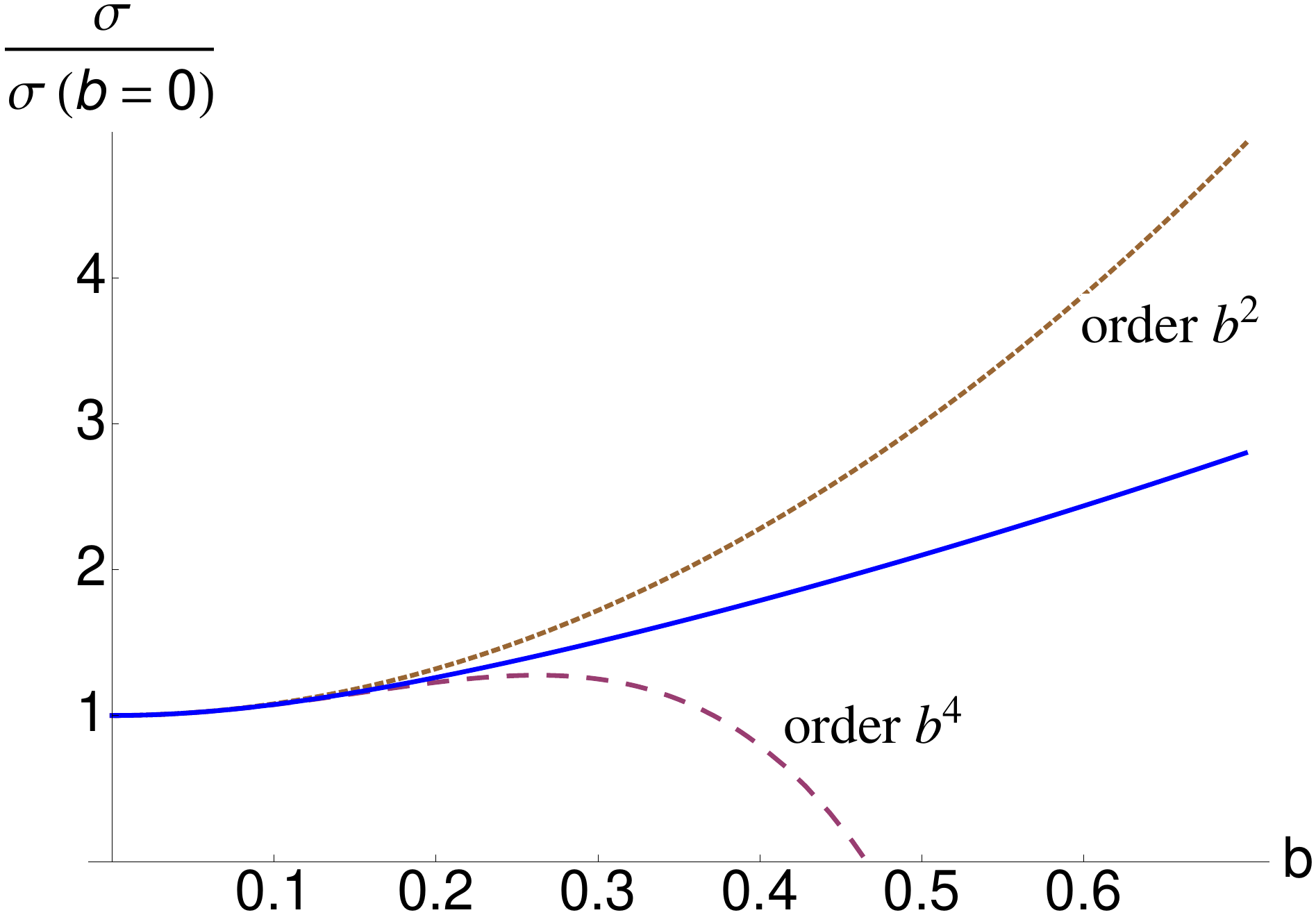} 
\caption{Longitudinal conductivity as a function of the parameter $b$. Solid line is the exact calculation, dotted line 
is the 4th order expansion in b calculated through the Boltzmann equation, dashed line is the 2nd order.}
\label{fig:sigma_b}
\end{figure}

The Diffuson structure factor $\Gamma^{(d)}$ in Eq.(\ref{eq:sigmaB}) is defined diagrammatically in 
Fig.~\ref{fig:DiffusonPropag}, and 
satisfies the usual recursive Dyson equation, solved by the expression 
 $\Gamma^{(d)}(\vec{q}, \omega) = \gamma \left[ \mathbb{1} \otimes \mathbb{1} - 
 \gamma P^{(d)}(\vec{q}, \omega) \right]$ 
where $P^{(d)}$ is the polarizability 
\begin{equation}
P^{(d)}(\vec{q},\omega) =  \int \frac{d\vec{k}}{(2 \pi)^2} 
\langle G^{R}(\vec{k},E) \rangle \otimes \langle G^{A}(\vec{k}-\vec{q},E -\omega) \rangle, 
\end{equation}
where $\otimes$ denotes a tensor product of the spin Hilbert spaces. 
 The resulting structure factor $\Gamma^{(d)}$ is naturally decomposed into 4 spin modes. However, as opposed to 
 the case of non-relativistic 
 electrons \cite{Akkermans:2007}, 
 for finite but small $q$, the non-diagonal character of the Dirac Green's functions leads to unusual terms   in this structure factor 
 such as 
 $(\vec{q}.\vec{\sigma})\otimes (\vec{q}.\vec{\sigma}), (\vec{q}.\vec{\sigma})\otimes \mathbb{1} $. 
 We recover the standard singlet and triplets states only in the diffusive $q\to 0$ limit. 
 In this limit, the only gapless mode is the singlet state, characteristic of the
  symplectic / AII class. 
 While this mode and the associated Cooperon singlet determines the quantum corrections to diffusion, it does not 
 contribute to 
 the classical conductivity of Eq.~(\ref{eq:sigmaB}).  The only classical contribution comes from one of the massive 
 triplet states. 
 In full generality, the $q\to 0$ limit of the structure factor $\Gamma^{(d)}$ can be parameterized according to 
 \begin{equation}
 \Gamma^{(d)} 
 = a_1 \mathbb{1} \otimes \mathbb{1} + a_2 (\sigma^x \otimes \sigma^x + \sigma^y \otimes \sigma^y) + a_3 
 \sigma^z \otimes \sigma^z . 
 \end{equation}
Using the parameterization of Eq.~(\ref{eq:J}) and performing the resulting trace we obtain :
\begin{equation}
\sigma_{B} = j^2 (a_1 -a_3) = \frac{\hbar e^2 v_F^2}{2 \pi \gamma} \frac{(\alpha(b) 
+ 2 \beta(b))^2}{\alpha(b)(\alpha(b)+\beta(b))}
\label{eq:sigma_B}
\end{equation} 
Adding the two contributions of Eq. (\ref{eq:sigma_A}) and (\ref{eq:sigma_B}) leads to the full expression for the 
classical conductivity : 
\begin{equation}
\sigma_{cl} = \sigma_{A} + \sigma_{B} =  e^{2}\rho D.
\end{equation}
The evolution of this classical conductivity as a function of the warping amplitude is represented in Fig. \ref{fig:sigma_b}.
Moreover, this expression of the conductivity allows us to define the diffusion constant $D$ from the Einstein relation,  
whose final expression reads 
\begin{eqnarray}
\label{eq:ExactD}
D &=& 
\frac{v_F^2 \tau_{e}}{2}
\left( 
\frac{\alpha(b) +5 \beta(b) + \delta(b)}{\alpha(b)}
+
\frac{(\alpha(b) + 2 \beta(b))^2}{\alpha(b)(\alpha(b)+\beta(b))}
\right)
\end{eqnarray}
which is non perturbative in the warping parameter $b$.
\begin{figure} [!h]
\centering 
\includegraphics[width=10cm]{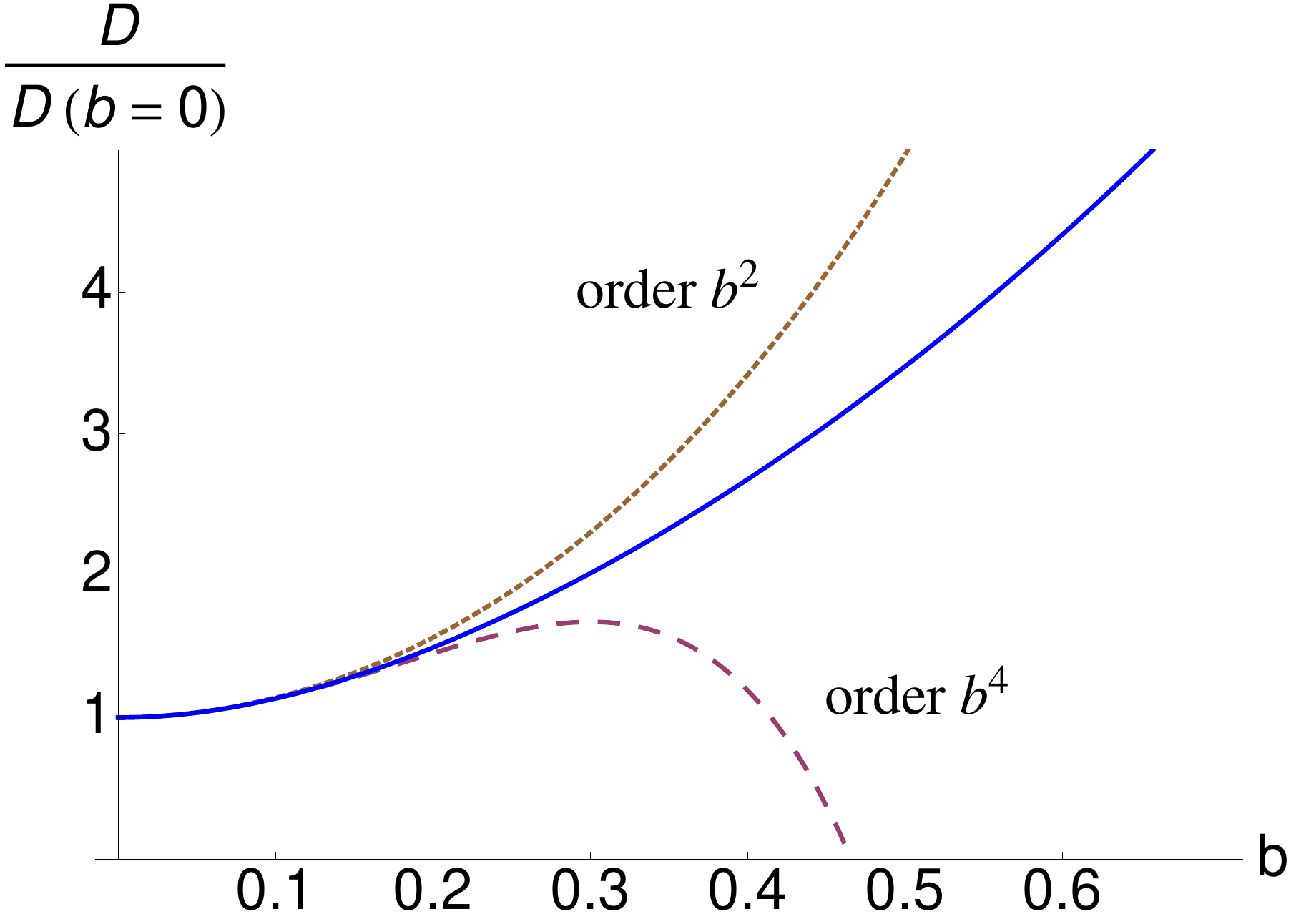} 
\caption{Evolution of the diffusion constant as a function of warping intensity $b$. The solid line corresponds to the expression 
non perturbative in the warping amplitude $b$, while the dotted and 
dashed curves correspond to perturbative results to order $b^{2}$ 
and $b^{4}$. }
\label{fig:diffusion_constant}
\end{figure} 
 In the limit $b \rightarrow 0$ of absence of warping, 
we recover the known result for Dirac fermions in the presence of a scalar disorder with a diffusion constant 
$D = v_F^2 \tau_e = v_F^2 \tau_{tr}/2 $ 
and  a transport time $\tau_{tr} = 2 \tau_e$ accounting for the inherent anisotropic scattering of Dirac fermions on 
scalar disorder.  
 The renormalized diffusion constant is found to increase as a function of $b$ (Fig. \ref{fig:diffusion_constant}). This 
 dramatic 
increase of $D$ within the range of experimentally relevant warping parameter $b$ signals an effective strong 
increase of the anisotropy 
of scattering when hexagonal warping is present. Note that in the present case, this diffusion coefficient accounts for 
both the renormalization of the averaged Fermi velocity and the scattering amplitude corresponding to a 
renormalized transport time. 
Moreover, the comparison of the expression of Eq.~(\ref{eq:ExactD}) with its perturbative expansion to order $b^4$ 
demonstrates that 
the experimental values of warping $b$ are beyond the reach of this perturbative expansion (and the
 similar perturbative studies of \cite{Tkachov:2011,Wang:2011})
: to describe accurately the experimental situation at high doping, the full exact expression Eq.~(\ref{eq:ExactD})  is 
required.

\section{Quantum corrections to the conductivity} 
\label{sec:quantum}

In this section, we focus on the regime of quantum transport corresponding to a phase coherent diffusion of the 
surface states 
which can be reached for small sample size and at low temperatures.  This regime corresponds to the situation 
where the Fermi momentum $k_{F}$ satisfies the condition $k_{F}l_{e}\gg 1$, which naturally corresponds to 
the situation where the warping of the Fermi surface is strong. 
 We focus on the  first two cumulants of the distribution of conductivity and compute diagrammatically the 
 corresponding quantum corrections non perturbatively in the warping amplitude $b$,  extending the results 
 of Refs. \cite{Tkachov:2011,Wang:2011}.

\subsection{Universality classes}  

The study of non-interacting metals perturbed by disorder corresponds to the well-studied problem of 
Anderson localization of electronic waves. In this framework, the transport properties depend on both 
dimension and universality 
classes determined by the symmetries preserved by disorder. These universality classes encodes both the 
universal properties of the Anderson transition, but also the universal properties of the weak disorder 
metallic regime of interest experimentally in the present case. 
The model considered in this paper $H = \hbar v_F \vec{\sigma} \cdot \vec{k} + V(\vec{r})$ 
is described at large distances by the standard AII / symplectic class : 
 time-reversal symmetry $T$ is preserved, but due to the momentum-spin locking it squares to $T^{2}=-\mathrm{Id}$ 
 \cite{Altland:1997,Evers:2008}. 
 This is also the Anderson universality class describing non-relativistic electrons in the presence of random spin orbit disorder. However while both 
 models possess  the same universal weak localization properties, they differ in their strong disorder behavior : the topological nature of the 
 single Dirac species in the present model manifest itself in the appearance of a topological term in the 
 field-theoretic description of the $d=2$ AII class \cite{Schnyder:2008,Schnyder:2009}. The presence of this 
 topological $\mathbb{Z}_{2}$ term modifies the strong disorder / low conductance behavior of the AII 
 class \cite{Bardarson:2007,Nomura:2007,Ostrovsky:2010}, but plays not role in the weak disorder regime. 
 Hence the behaviors of the first two cumulants in the weak localization regime of the single Dirac model with scalar disorder are exactly 
 those of the standard AII / symplectic class. 
  Moreover adding the HW term $H_w$ does not break Time Reversal Symmetry, and thus does not change the universal symmetry class. 
In contrast,  either a magnetic disorder or a magnetic field introduced  through a Zeeman or orbital  term in the Hamiltonian
 breaks time-reversal symmetry and induces 
 a crossover from the AII/symplectic to the A/unitary class characterized by the absence of time reversal symmetry. 
Such a crossover between the symplectic and unitary ensembles has already 
been observed by depositing Fe impurities at the surface of ultrathin samples of Bi$_2$Te$_3$ \cite{He:2011}. In this 
experiment, the exchange field of the Fe atoms induces the Zeeman coupling for the SS carriers and destroys the weak 
antilocalization signature. Nevertheless the extreme thickness of the sample is likely to yield an important 
hybridization between the bulk, and  top/bottom SSs while here we consider the response of a single isolated SS.
 
 One of the  characterizations of Anderson universality symmetry classes is the number of independent low energy
modes in the diffusive regime. 
Those modes correspond to the various Cooperon/Diffuson modes which indeed parametrize the target space of the underlying non-linear sigma field theory. 
The symplectic class corresponds to two modes, one singlet  Diffuson mode and one singlet Cooperon mode \cite{Akkermans:2007}. 
 The breaking of time reversal symmetry by {\it e.g.} an 
 orbital magnetic field will suppress the Cooperons whose propagator acquires a mass and is no longer diffusive : 
 this induces  a cross-over from the symplectic class to the unitary class, characterized by a single diffusive mode. This cross-over 
 occurs on a length-scale determined by the mass acquired by the Cooperon singlet. 
 The suppression of the Cooperon is attributed to destructive interference between time-reversed paths, accounted for by the retarded and advanced Green's functions in the Kubo formula. Such a cross-over will be discussed in section \ref{sec:Zeeman}.

\subsection{Weak antilocalization}
When winding around the Fermi surface, the spinorial electron wave function acquires a $\pi$ Berry phase which is 
responsible for weak antilocalization (WAL) phenomenon. In graphene, the presence of intervalley scattering leads 
to a crossover from weak antilocalization to weak localization when the ratio of intervalley over intravalley scattering 
rates is increased \cite{McCann:2006,Tikhonenko:2009}. In contrast, TIs with a single Dirac cone provide ideal 
systems to measure WAL once the issue of the spurious bulk conductance is solved. Actually some experiments 
have already reported a strong WAL signal in thin films and in 3D HgTe slabs \cite{Bouvier:2011}.

 The quantum correction to the conductivity in the class AII corresponds to a weak anti localization correction : its determination within diagrammatic theory for Dirac fermions is recalled in  \ref{app:WAL}. In the limit of a  phase coherent sample of size 
 $L_{\phi}\ll L$ where  $L_{\phi}$ is the phase coherent length, it reads:
\begin{eqnarray}
\langle \delta \sigma \rangle    
&=& \left( \frac{e^2}{\pi \hbar} \right)
\int_{\vec{Q}}\frac{1}{Q^2}
= \frac {e^2}{\pi h} \ln (L_{\phi}/\ell_e) 
\end{eqnarray}  
where $\ell_e$ is the elastic mean free path. This result is independent of the particular model within the class AII. However, it depends on the 
 the amplitude $b$ of warping at high Fermi energy through the
 diffusion coefficient entering the phase coherence length $L_{\phi} =
 \sqrt{D(b) \tau_{\phi}}$,  where $\tau_{\phi}$ is the phase coherence time. 

When a transverse magnetic field is applied, the standard derivation to account for the magnetic orbital effect of this result still holds
\cite{Akkermans:2007}: it amounts to derive the Cooperon contribution from the probability of return to the 
origin of a diffusive path 
in the presence of the magnetic field. 
The corresponding correction to the conductivity is thus given by \cite{Akkermans:2007}:
\begin{equation}
\langle \delta \sigma (B) \rangle = 
\frac{e^2}{4 \pi^2 \hbar} \left[
\Psi \left(\frac{1}{2}+\frac{B_{e}}{B} \right) 
- \Psi \left(\frac{1}{2}+\frac{B_{\phi}}{B}\right) \right], 
\end{equation}
where the characteristic fields $B_{e} = \hbar /(4e D(b) \tau_{e})$ and $B_{\phi} = \hbar /(4e D(b) \tau_{\phi})$ have been introduced, and where $\Psi$ is the Digamma function. The diffusion constant $D(b)$, non perturbative in $b$, is given by Eq.(\ref{eq:ExactD}).
This corresponds to the result obtained for graphene when inter-valley scattering can be neglected \cite{McCann:2006}. 
In our case, this expression 
implies that in a given sample, the shape of the weak anti-localization correction as a function of $B$ will evolve as the Fermi energy is varied within the Topological Insulator gap and the warping amplitude $b$ is varied. This effect will be discussed further in 
section \ref{sec:Results}.

\subsection{Universal Conductance Fluctuations}

As explained above, in the absence of magnetic field the Universal Conductance Fluctuations (UCF) result for
 the symplectic class results still holds and one finds, similarly to graphene \cite{Kharitonov:2008}~:
\begin{equation}\label{eq:UCF-Dirac}
\langle \delta \sigma^2 \rangle = 
 12 \left( \frac{e^2}{h} \right)^2 \  \frac{1}{V}\! \int_{\vec{q}}\frac{1}{q^4} . 
\end{equation}
The derivation of this result for Dirac fermions is recalled in \ref{app:UCF}. 
Defining the phase coherent  $L_{\phi} = \sqrt{D(b) \tau_{\phi}(T)}$ and thermal lengthscales 
$L_{T} = \sqrt{\hbar D(b) / T}$, we can focus on different regimes : 
 for $L \ll L_{\phi},L_{T}$, a proper regularization of the integral in (\ref{eq:UCF-Dirac}) leads to the universal value 
\begin{equation}\label{eq:UCF-Dirac-1}
\langle \delta \sigma^2 \rangle = 
\frac{12}{\pi^{4}} \left( \frac{e^2}{h} \right)^2
\sum_{n_{x}=1}^{\infty} \sum_{n_{y}=0}^{\infty}
\frac{1}{(n_{x}^{2}+n_{y}^{2})^{2}} 
\simeq  0.185613 \left( \frac{e^2}{h} \right)^2 . 
\end{equation}
This result is independent of the diffusion coefficient $D(b)$ and thus independent of the warping amplitude. On the other hand, in the 
other limits \cite{Akkermans:2007}:
\begin{eqnarray}
\label{eq:UCF-Dirac-2}
\langle \delta \sigma^2 \rangle & \simeq & 
\frac{3}{\pi} \left( \frac{e^2}{h} \right)^2
\left( \frac{L_{\phi}}{L} \right)^{2}
\textrm{ for }L_{\phi} \ll L,L_{T} , 
\\
& \simeq & 
\left( \frac{L_{T}}{L} \right)^{2}\left( \frac{e^2}{h} \right)^2
\ln \left( \frac{L}{L_{T}} \right)
\textrm{ for }L_{T}\ll L \ll L_{\phi}   , 
\\
& \simeq & 
\left( \frac{L_{T}}{L} \right)^{2}\left( \frac{e^2}{h} \right)^2
\ln \left( \frac{L_{\phi}}{L_{T}} \right)
\textrm{ for }L_{T}\ll L_{\phi} \ll L . 
\end{eqnarray}
Hence a strong dependence of these UCF on the Fermi energy through the warping amplitude $b$ is found in all these cases. 
%
%
%
The introduction of a transverse magnetic field induces a crossover from the symplectic class to the unitary class, where the amplitude of the fluctuations is reduced by a factor of two. This crossover is described as \cite{Akkermans:2007}~:
\begin{equation}
\langle \delta \sigma(B)^2 \rangle = \frac{1}{2} \langle \delta \sigma^2 \rangle \left[ 1 + \frac{B_{\phi}}{B} \Psi ' (\frac{1}{2} + \frac{B_{\phi}}{B}) \right]
\end{equation}
where $\Psi$ is the Digamma function and $B_{\phi}(b) = \hbar / 4eD(b) \tau_{\phi}$. The dependence in $b$ of 
the diffusion constant will affect the value of $B_{\phi}$ :  this characteristic field for the suppression of the WAL 
correction 
and this reduction by a factor $2$ of the conductance fluctuations 
will decrease when the Fermi level is raised away from the Dirac point.

\subsection{In-plane magnetic field: interplay between Zeeman  and warping effects}
\label{sec:Zeeman}

We now consider the effect of an in plane Zeeman field on the transport properties of these Dirac fermions. This can 
be accounted for by adding to the Hamiltonian  (\ref{eq:Dirac}) a term 
\begin{equation}
\label{eq:H-Zeeman}
H_{Z} = g \mu_{B} \left( \sigma_{x} B_{x} + \sigma_{y} B_{y}\right). 
\end{equation}
Without warping, such a Zeeman field acts exactly like a constant vector potential and can be gauged away. This 
amounts to shift uniformly the momenta by $-g \mu_B \vec{B} / (\hbar v_{F})$. Hence without warping, an in-plane 
Zeeman field does not modify the scattering properties of an (infinite) disordered sample. 
However the presence of the hexagonal warping term in (\ref{eq:Dirac}) breaks this invariance : the shape of the 
Fermi surface is now modified by the Zeeman field. As a consequence, the scattering amplitudes 
$f(\theta, \theta'-\theta)$ acquire also a $\vec{B}$ dependence. We can naturally expect that this Zeeman field 
which modifies the scattering amplitudes redistributes the scattering matrix of the samples, 
and leads to  conductance fluctuations induced by an in-plane magnetic field. 

\begin{figure} [!h]
\centering 
\includegraphics[width=10cm]{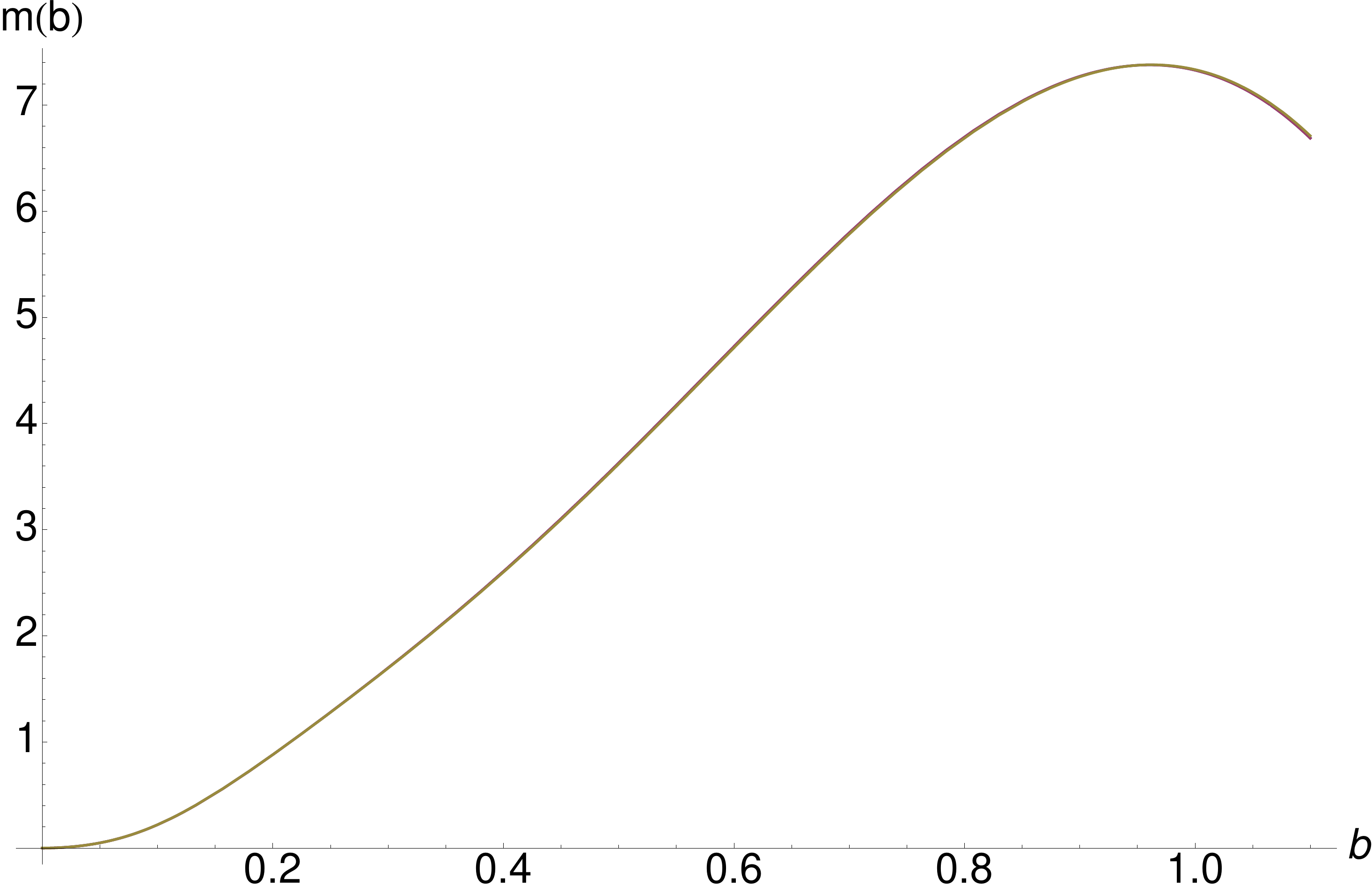} 
\caption{Evolution of the massive term $m(b)$ in the diffusion equation of the Cooperon, in presence of the 
hexagonal warping and in-plane Zeeman magnetic field, as a function of the warping parameter b.}
\label{fig:zeeman+warping mass}
\end{figure}

 To describe quantitatively the effect of this Zeeman field, we extend the above diagrammatic analysis perturbatively
  in $\tilde{B}=g \mu_B B/ E_F$ . 
This Zeeman field, which breaks time reversal symmetry, induces a crossover from the symplectic to the unitary class.
This crossover can be accounted for by the evolution of the Cooperon structure factor : its singular part does not 
correspond anymore to a diffusive 
singlet component 
but to the diffusion of 
a massive singlet:
\begin{equation}
 \Gamma^C(\vec{Q}) = \frac{\gamma}{Dq^2 \tau_e + m(b,\tilde{B})} \vert S \rangle \langle S \vert . 
 \end{equation}
The mass $m(b,\tilde{B})$ which encodes the effects of the Zeeman field, is calculated perturbatively to the second 
order in $\tilde{B}$ : 
$m(b,\tilde{B})=m(b)\tilde{B}^2$. The non-perturbative result is represented in Fig. \ref{fig:zeeman+warping mass}. 
We recover the expected result $m(0,\tilde{B})=m(b,0)=0$ corresponding to the 
single singlet diffusive Cooperon mode of the symplectic class. 
%
This mass is  associated with a 
length parameterizing the cross-over  from the symplectic to the unitary class, defined by 
$L_{B}=\sqrt{D \tau_e /m(b,B)}$. Beyond this length scale, we recover a standard unitary weak localization, and 
conductance fluctuations are reduced by a factor $2$ : 
\begin{eqnarray}
\langle \delta \sigma (\vec{B}) \rangle    
 &= \left( \frac{e^2}{\pi \hbar} \right)  \int_{\vec{Q}}\frac{1}{Q^2 + L_{B}^{-2}+L_{\phi}^{-2}} 
 = f(\tilde{L}/\ell_e,L/\tilde{L}) 
 \\
\langle \delta \sigma^{2} (\vec{B}) \rangle &=
6  \left( \frac{e^2}{h} \right)^2 \  \frac{1}{V}\! 
\left[ \int_{\vec{q}}\frac{1}{(q^2+L_{\phi}^{-2})^2}
+ \int_{\vec{q}}\frac{1}{(q^2+L_{B}^{-2}+L_{\phi}^{-2})^{2}}
\right] \nonumber \\
& = \frac12 \langle \delta \sigma^{2} (\vec{B}=\vec{0}) \rangle + f_{2}(L/\tilde{L}), 
\end{eqnarray} 
where $L$ is the longitudinal size of the topological insulator surface, $L_{\phi}$ is the phase coherence length and 
$\tilde{L}$ is defined as $\tilde{L}^{-2} = L_{B}^{-2}+L_{\phi}^{-2}$. 
The function $f$ and $f_2$ depends on the geometry of the sample. In the case of an infinite 2D sample, one 
recovers only a logarithmic dependence on $\tilde{L}/\ell_e$ for $f$. For a finite size sample, one has to to replace 
the integral by a sum over the compatible wavevectors\cite{Akkermans:2007}.

\begin{figure} [!h]
\centering 
\includegraphics[width=10cm]{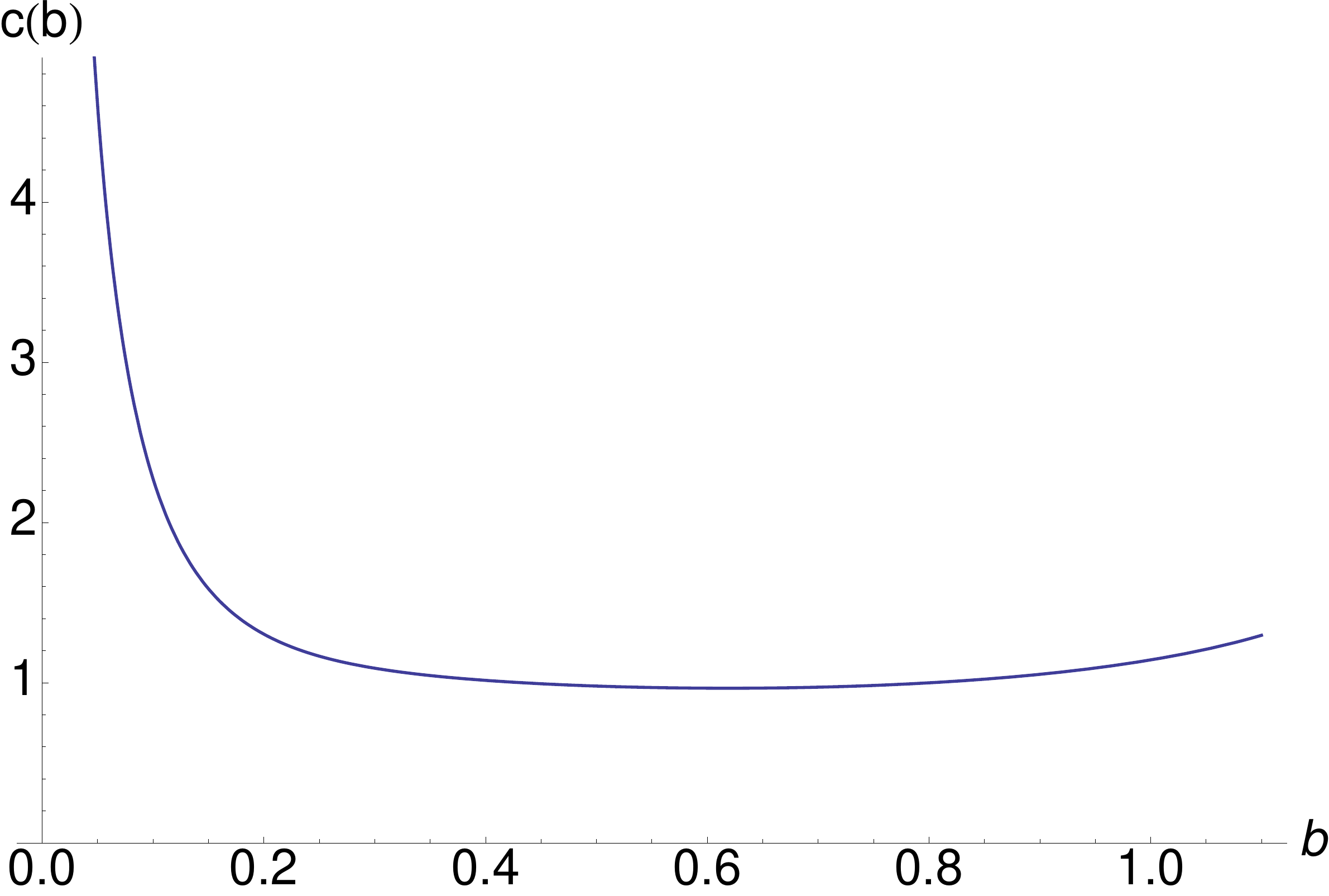} 
\caption{Dependence in $b$ of the characteristic length to observe the crossover from symplectic to unitary class, 
for an in-plane Zeeman magnetic field.}
\label{fig:zeeman+warping mag_length}
\end{figure}

As expected, we find that the UCF extrapolates between the value for the symplectic and the unitary class, over the 
length scale $L_{B}(b,B)$. An estimation of the evaluation of this length as a function of the magnetic field or the 
warping term is given by :
\begin{equation}
\frac{L_{B}}{\ell_e} = \frac{\sqrt{D(b)/D(b=0)m(b)}}{\tilde{B}} = \frac{c(b)}{\tilde{B}}, 
\end{equation}
where the evolution $c(b)$ with the warping amplitude is shown in Fig. \ref{fig:zeeman+warping mag_length}
For experimental realistic values, $b\simeq 0.7$, $c(0.7) \simeq 1 $, $E_F \simeq 0.3 \quad eV$ and $g \mu_B 
\simeq 5.10^{-4} \quad eV.T^{-1}$. This gives for a magnetic field around 1 T a characteristic length to observe 
the crossover $L_B \simeq 1000 \ \ell_e$ which is is rather large but within experimental reach.


\section{Summary of Results and Conclusions}
\label{sec:Results}

 In this section, we summarize the main results of our approach focusing on the experimentally
 relevant aspects. In particular, we emphasize the effect of strong variation of the warping
 amplitude $b$ as the Fermi energy is varied in a given material. 
Universal properties of  transport in the quantum regime 
are not affected by the presence of this warping.
However, both the departure from these universal properties as well as the classical regime 
are entirely characterized  by the diffusion constant $D(b)$ which itself depends
in the warping amplitude.

\begin{figure} [!h]
\centering 
\includegraphics[width=10cm]{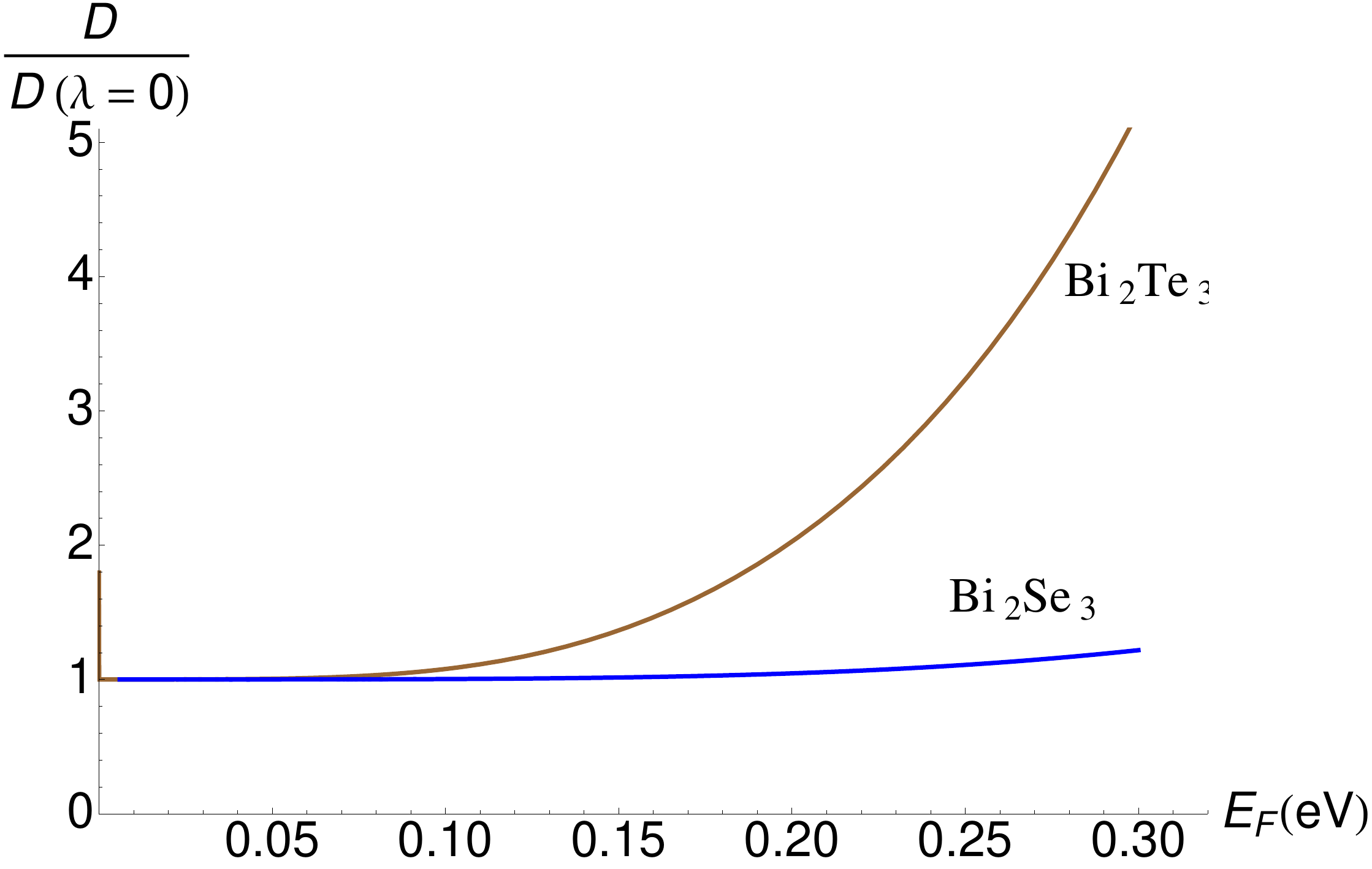} 
\caption{Dependence of the diffusion constant on the Fermi energy for the  values of 
$\lambda,v_{F}$ 
corresponding to the Bi$_{2}$Se$_{3}$ and  Bi$_{2}$Te$_{3}$ compounds. The results depend on the amplitude of 
disorder 
parameterized by the mean free path $l_{e}$. Here, they  are normalized with respect to the diffusion constant in 
absence of warping 
$D(\lambda=0)$ which is independent of the Fermi energy and incorporates the dependence on $l_{e}$.}
\label{fig:D_E}
\end{figure}

\begin{figure} [!h]
\centerline{
\includegraphics[width=8cm]{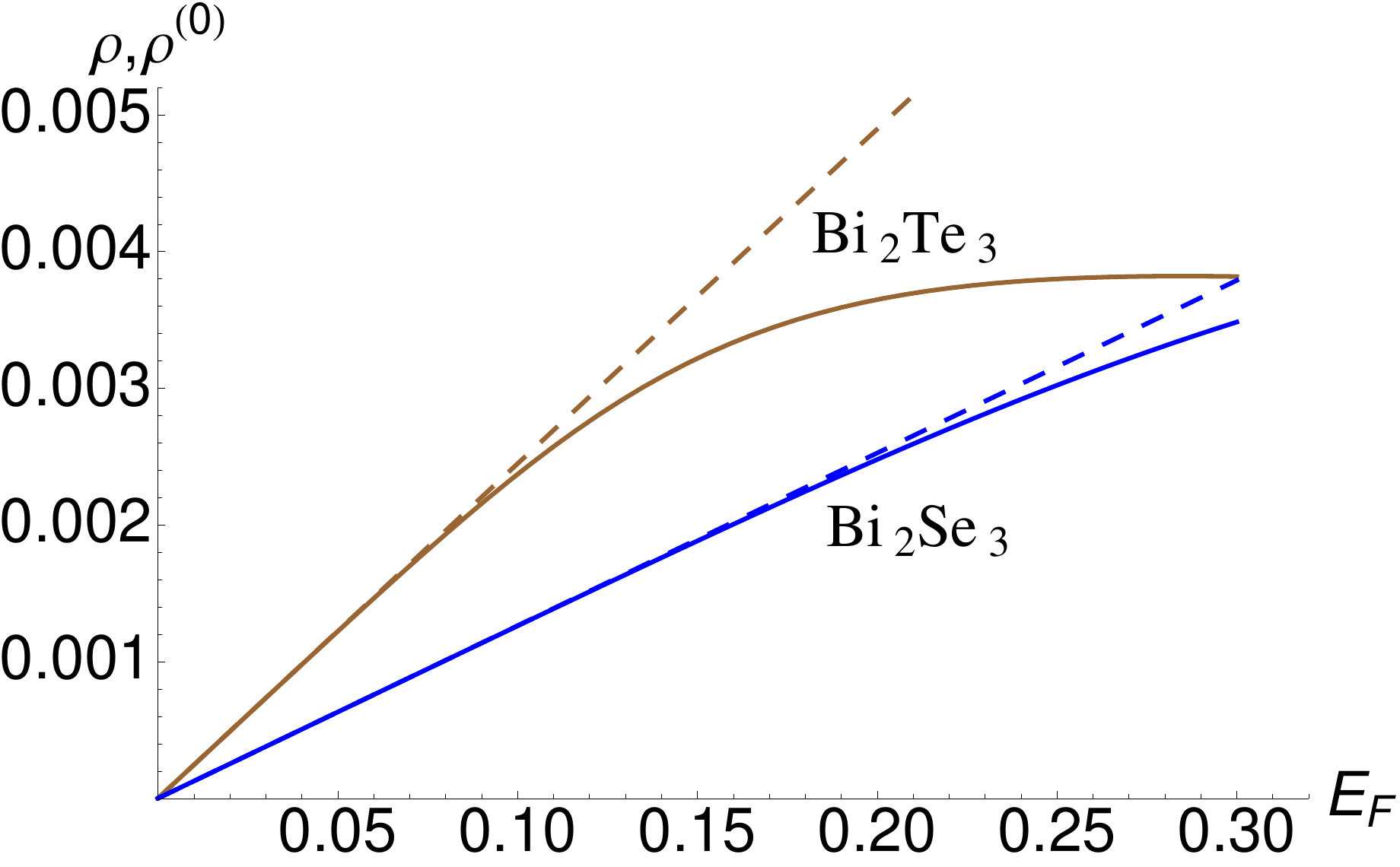} 
\includegraphics[width=8cm]{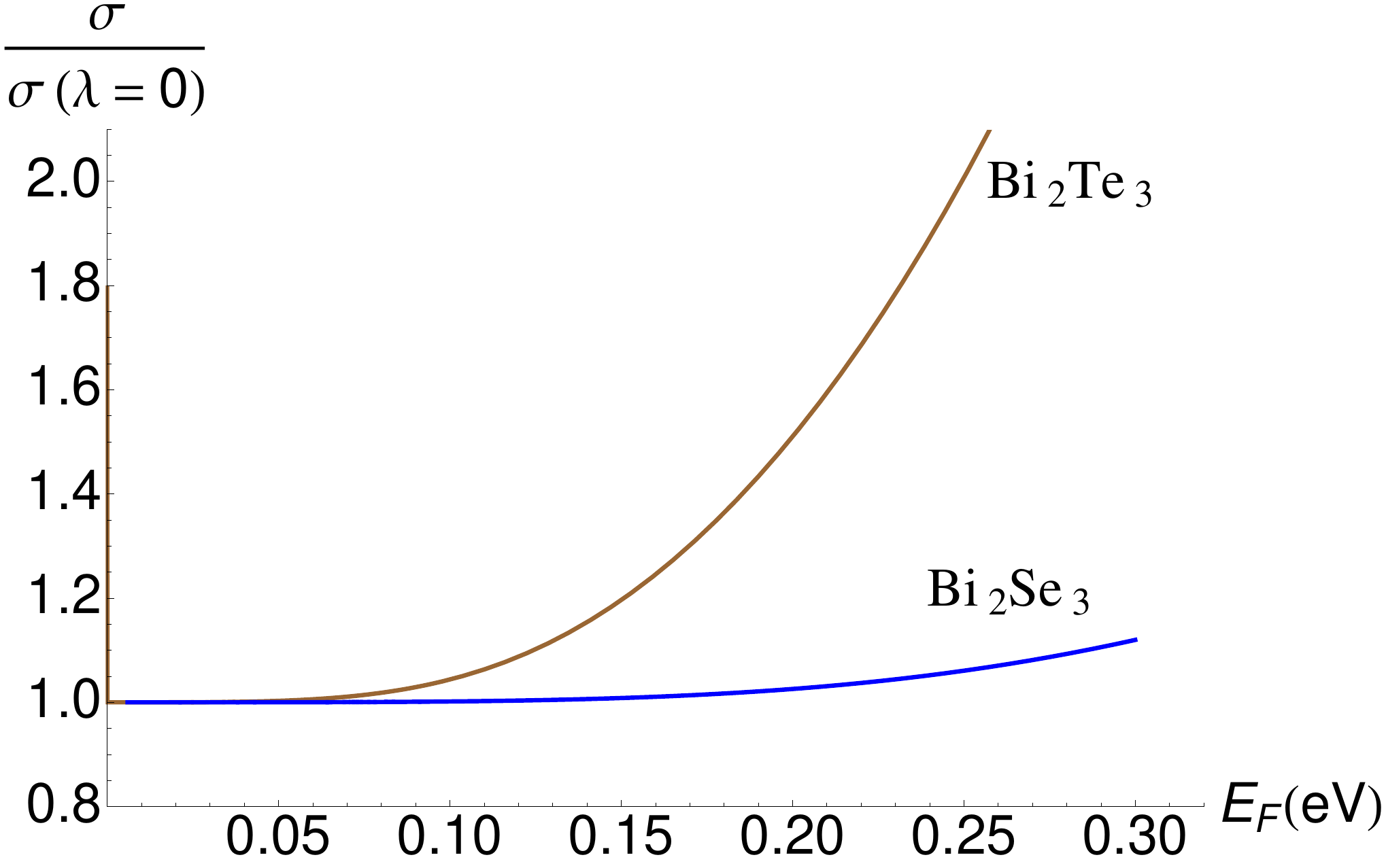} 
}
\caption{
Dependence of the density of states  (Left) and 
classical conductivity (Right) on the Fermi energy 
for values of 
$\lambda = 128\ \textrm{eV}.\mathring{A}^3$ , $v_F=3.55 \ \textrm{eV}.\mathring{A}$ for 
 Bi$_{2}$Se$_{3}$  and 
$\lambda = 250\ \textrm{eV}.\mathring{A}^3$ and $v_F=2.55 \ \textrm{eV}.\mathring{A}$
corresponding to the  Bi$_{2}$Te$_{3}$ compounds.
 On the left figure, 
 the dotted lines corresponds to the standard linear density of states for Dirac fermions
  without warping  $\lambda =0$.  
On the right figure, the results are represented as a ratio with 
the 
conductivity in the absence of warping $\sigma(\lambda =0)$ which is independent on the energy and incorporates 
the dependence 
on the disorder strength. 
}
\label{fig:densitystates}
\end{figure}

 The transport properties in the incoherent classical regime are characterized by both the density of states 
 $\rho(E_F)$ and the 
 diffusion constant $D$. 
The dependence on the Fermi energy of both quantities is strongly affected
  by the presence  of the warping of the Fermi surface, whose amplitude $b$ itself depends on $E_F$. 
  This is shown on
  Fig.~\ref{fig:D_E},\ref{fig:densitystates} for two sets of parameters
  corresponding to the Bi$_{2}$Se$_{3}$ and  Bi$_{2}$Te$_{3}$ compounds. As a
  consequence of these results, the conductivity $\sigma$ acquires a strong
  Fermi energy dependence shown on Fig.~\ref{fig:densitystates}, which can be
  directly probed experimentally. 
%

\begin{figure} [!h]
\centerline{
\includegraphics[width=8cm]{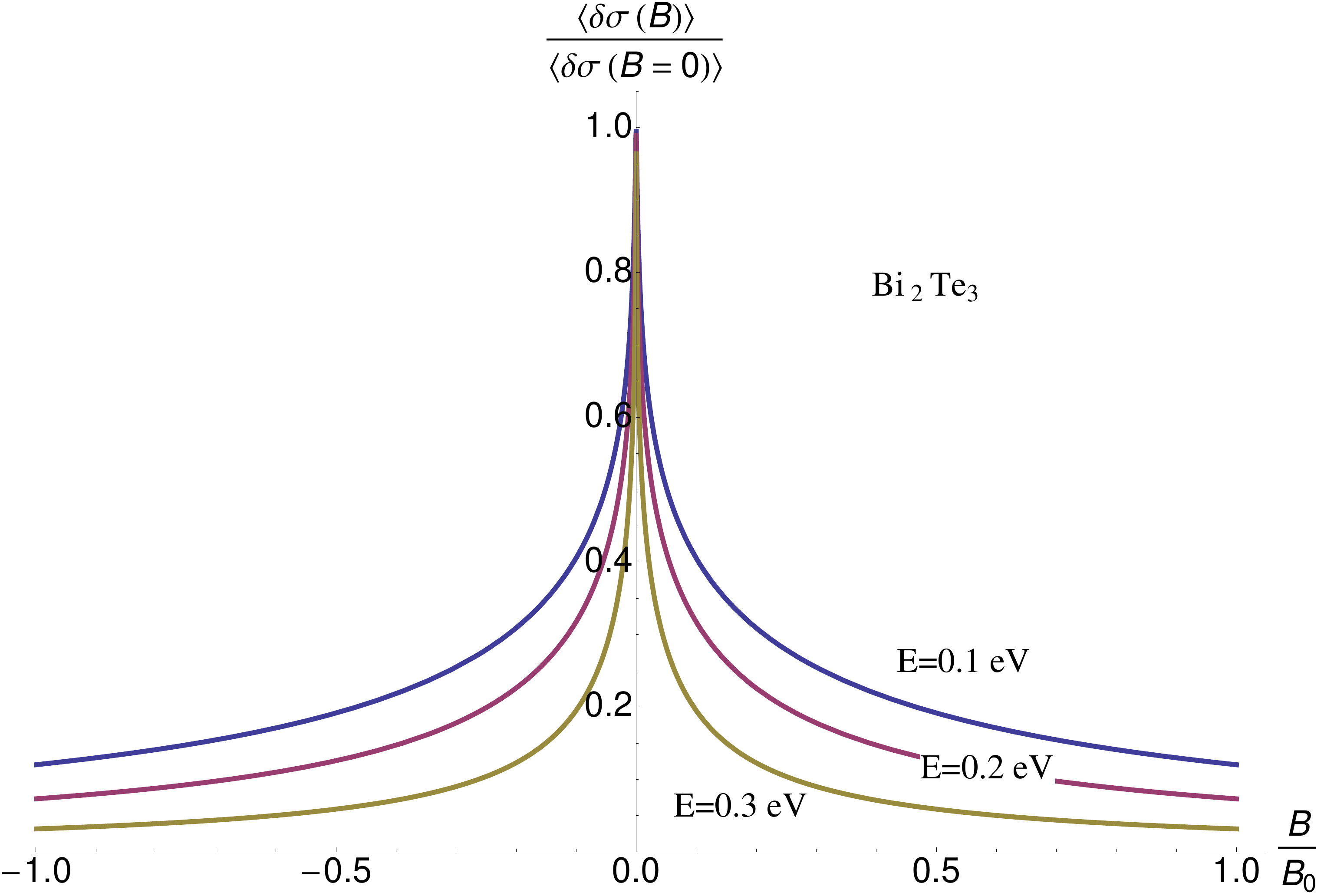} 
\includegraphics[width=8cm]{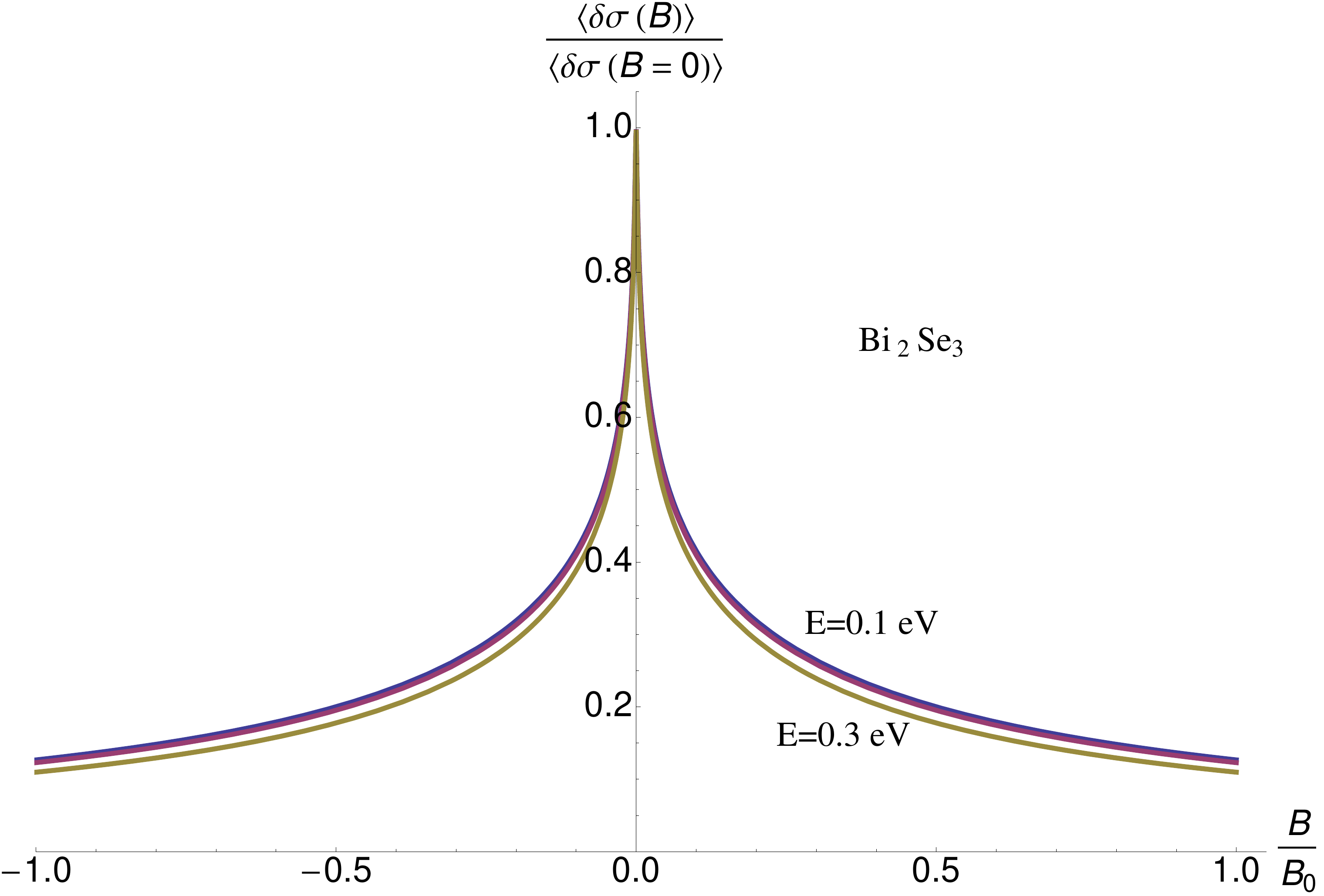} 
}
\caption{Dependence of the weak localization correction 
$\langle \delta \sigma (B) \rangle /  \langle \delta \sigma (B=0) \rangle$
on the Fermi energy for the values of $\lambda,v_{F}$ 
corresponding to the  Bi$_{2}$Te$_{3}$  (Left) and Bi$_{2}$Se$_{3}$ (Right) compounds. 
We have chosen to scale the magnetic field as $B/B_{0}$ where $B_{0} l_{e}^{2} = \phi_{0}=h/e$ to avoid any energy 
(or warping) dependence 
of this rescaling field. The results show a clear dependence on energy 
of the magnetic field characteristic of  weak localization decay. 
}
\label{fig:locfaible_B}
\end{figure}

\begin{figure} [!h]
\centering 
\includegraphics[width=10cm]{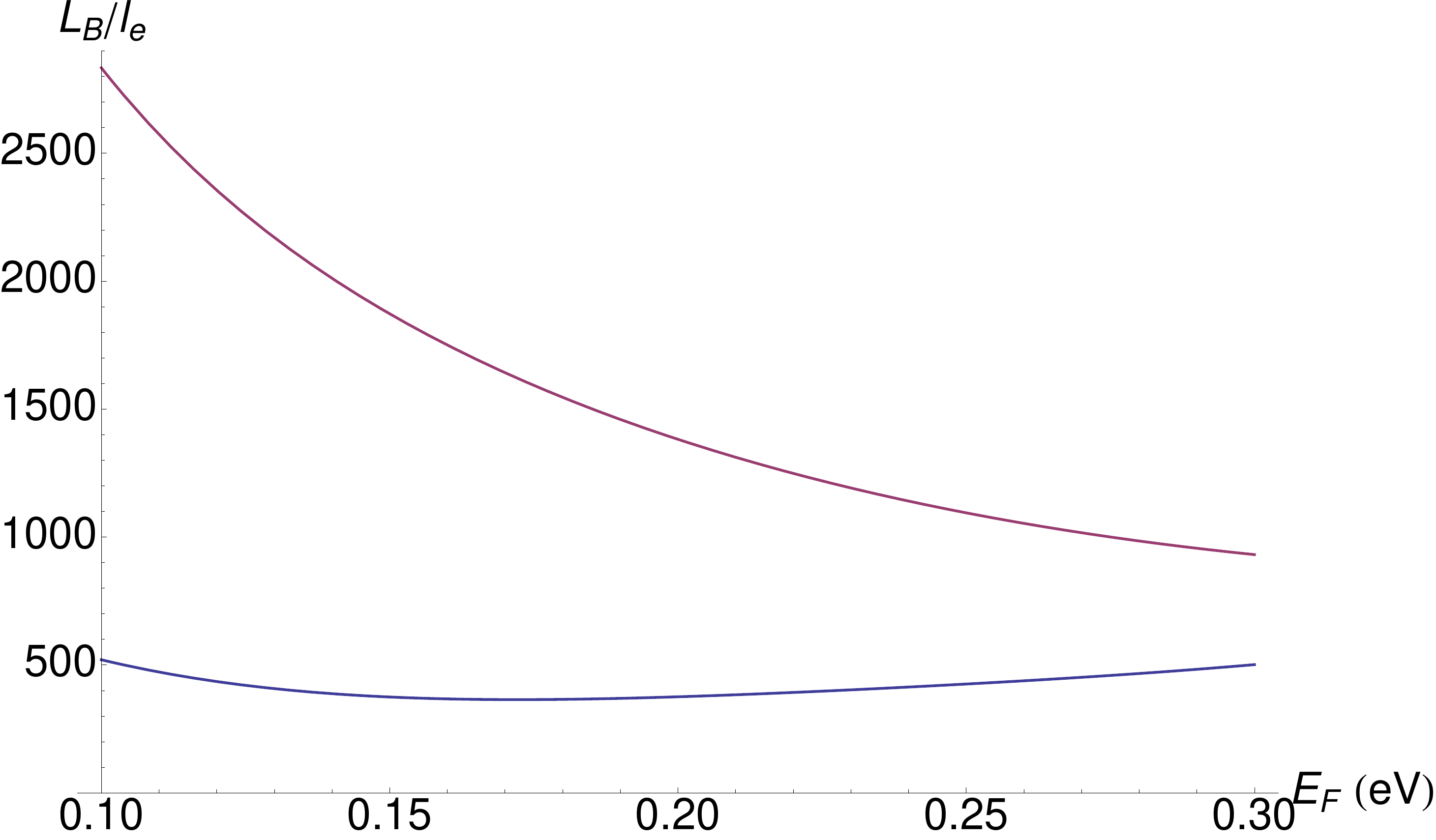} 
\caption{Dependence of the magnetic dephasing length $L_B$ on the Fermi energy for parameters corresponding 
to   Bi$_{2}$Te$_{3}$  (Red) and Bi$_{2}$Se$_{3}$  compounds.
}
\label{fig:warping_zeeman_l}
\end{figure}

Similarly, the quantum corrections to transport depends strongly on the Fermi
energy through the dependence of the Diffusion coefficient $D$ on the warping
amplitude. Indeed, the shape of the typical measurement of the weak (anti-)localization 
correction through the dependence of the conductivity on a magnetic field
perpendicular to the surface depends solely on the diffusion constant. This
diffusion constant depending on the Fermi energy, the associated
anti-localization curve depends itself on this Fermi energy as shown on
Fig.~\ref{fig:locfaible_B}. 
While the amplitude of conductance fluctuations are
universal in the limit of an entirely coherent conductor, their amplitude in a
realistic situation where $L_\phi \simeq L$ will be parameterized by a universal
function of the diffusion coefficient. We have shown moreover that these
fluctuations depend in a remarkable way on an in-plane Zeeman magnetic field.
The amplitude of this effect depends on the ratio between the associated magnetic
dephasing length $L_B$ and the mean free path $ \ell_e$. The dependence of this
ratio on the Fermi energy is shown on Fig.~ \ref{fig:warping_zeeman_l}  for the two same sets 
of values used above. 

In conclusion, we have shown that 
it is essential to 
take into account the hexagonal deformation of the Dirac cone
occurring at the surface  of topological insulators such as Bi$_2$Se$_3$ and Bi$_2$Te$_3$ 
to accurately describe both their classical and quantum transport properties. 
%
%
In particular, we provide a formula describing the evolution of diffusion constant $D$ 
for arbitrary strength of the warping. 
Since this warping amplitude increases with the Fermi energy of the surface states,   
we predict a dependance on doping of transport properties  different from 
those predicted for a perfect Dirac cone.

This work was supported by the ANR under the project 2010-BLANC-041902 (ISOTOP). JC also acknowledges support from EU/FP7 under contract TEMSSOC.

\begin{appendix}

\section{Non Perturbative Density of states}
\label{app:density}

In the presence of warping, the density of states at the Fermi level
$\rho(\epsilon_F)$ is modified. One has:
\begin{eqnarray}
 \label{eq:dos-def}
 \rho(\epsilon)=\int_0^{+\infty} \frac{k dk}{2\pi} \int_0^{2\pi}
 \frac{d\theta}{2\pi} 
 \delta \left[\epsilon - \sqrt{(\hbar vk)^2 + \lambda^2 k^6 \cos^2(3\theta)}\right].
\end{eqnarray}
Performing the angular integration first, the density of states is
obtained as the integral:
\begin{eqnarray}
 \label{eq:dos-single-int}
 \rho(\epsilon)=\frac{\epsilon}{2(\pi \hbar v)^2} \int_{y_-}^1
 \frac{dy} {\sqrt{(1-y)(4b^2 y^3 + y -1)}},
\end{eqnarray}
where $y_-$ is the unique real solution of the equation $4b^2 y^3 +
y-1=0$. By the usual formula for solving equations of the third
degree, we have:
\begin{eqnarray}
 \label{eq:yminus}
 y_-=\frac{1}{2\sqrt{3} b} \left[\sqrt[3]{3\sqrt{3} b +\sqrt{1+27
       b^2}} - \sqrt[3]{ \sqrt{1+27
       b^2} - 3\sqrt{3} b}\right].
\end{eqnarray}
It is convenient to introduce the parameterization $\sinh \varphi = 3
\sqrt{3} b$, giving $y_-=3\sinh (\varphi/3)/\sinh \varphi$. With the
change of variables $y_-=1-1/t$, the integral in
(\ref{eq:dos-single-int}) can be rewritten:
\begin{eqnarray}
 \label{eq:dos-single-int2}
  \rho(\epsilon)=\frac{\epsilon}{2 \sqrt{2} b (\pi \hbar
    v)^2}\int_{t_-}^{+\infty}\frac{dt}{\sqrt{(t-1)^3 -t^2/(2b^2)}}.
\end{eqnarray}
Using Eqs. (17.4.70)--(17.4.72) from \cite{Abramowitz:1972}, the
integral in Eq.~(\ref{eq:dos-single-int2}) can be expressed  in terms of a
complete elliptic integral of the first kind, giving the density of
states:
\begin{eqnarray}
 \label{eq:dos-final}
 \rho(\epsilon)=\frac{\epsilon}{2(\pi \hbar v)^2} \frac{2
   K\left[\frac 1 2 - \frac 1 4 \left(3 - \frac{1}{\left(1+\frac 4 3
           \sinh^2 (\varphi/3)\right)^2}\right) \sqrt{\frac{1+\frac
       4 3 \sinh^2 (\varphi/3)} {1+4\sinh^2
         (\varphi/3)}} \right]}{\left(1+\frac 4 3 \sinh^2
         (\varphi/3)\right)^{3/4} \left(1+ 4  \sinh^2
         (\varphi/3)\right)^{1/4}}
\end{eqnarray}
In the limit of $\lambda \to 0$, $\varphi \to 0$, and
(\ref{eq:dos-final}) reduces to the density of states in the absence
of warping. For large $\lambda$, we have $b \sim 2
e^{\varphi}{3\sqrt{3}}$, and the density of states behaves as:
$\rho(\epsilon) \sim 3^{-1/4} \pi^{-2} K(1/2-\sqrt{3}/4)(\epsilon
\lambda^2)^{-1/3}$. Since the density of states goes to zero for large
and small energy, it has a maximum at a finite value of $\epsilon$.
By simple scaling, the maximum is obtained for an energy $\epsilon^*=
C_1 \sqrt{(\hbar v_F)^3/\lambda}$ and $\rho(\epsilon^*)= C_2 (\hbar
v_F \lambda)^{-1/2}$. 

\section{Weak anti-localization correction for Dirac fermions}
\label{app:WAL}

In this appendix, we  derive the quantum correction to conductivity for two dimensional Dirac fermions 
in the absence of warping ($b=0$). In this case of linearly dispersing Dirac fermions, the current operator 
$j_x = -e v_F \sigma^x$ is no longer a function of $\vec{k}$. The renormalization of this current operator by 
vertex corrections can be written as 
\begin{equation} \label{eq:Vertex-direct}
\Sigma^{x} = j_{x} + j_{x}  P^{D} \Gamma^{D} = 2 j_x
\end{equation}
 where $\Sigma^{x}$ stands for the renormalized operator. 
The quantum corrections to conductivity are associated with contributions between interferences of loops of diffusive paths. 
This interference between a loop and its time-reversed contribution is described as the propagation of the so-called \emph{Cooperon}. 
Its propagator
 is defined through the Dyson equation
  $\Gamma^C(\vec{Q}, \omega) = \gamma [ \mathbb{1} \otimes \mathbb{1} - \gamma P^C(\vec{Q}, \omega)]$ where $P^C$ is :

\begin{equation}
P^{(C)}(\vec{Q},\omega) =  \int \frac{d\vec{k}}{(2 \pi)^2} 
\langle G^{R}(\vec{k},E) \rangle \ \langle G^{A}(\vec{Q}-\vec{k},E -\omega) \rangle . 
\end{equation}
By a time-reversal operation on the advanced component, we can relate  the Cooperon to the previously identified Diffuson propagator. 
Only the diffusive modes of the Cooperon contribute to the dominant quantum correction, leading to a  structure factor for the Cooperon : 
\begin{eqnarray}
\label{eq:Cooperon-propagator}
 \Gamma^C(\vec{Q}) &=& \frac{\gamma}{\tau_e} \frac{1}{DQ^2} \vert S \rangle \langle S \vert \\
 &=& \frac{\gamma}{\tau_e} \frac{1}{DQ^2} 
 \frac{ 1}{4} \left[\mathbb{1} \otimes \mathbb{1} - \sigma^x \otimes \sigma^x  - \sigma^y \otimes \sigma^y -  \sigma^z \otimes \sigma^z\right], 
 \end{eqnarray}
where D is the diffusion constant, $D = v_F ^2 \tau_e$ in the absence of warping.

\begin{figure} [!h]
\centering 
\includegraphics[width=7cm]{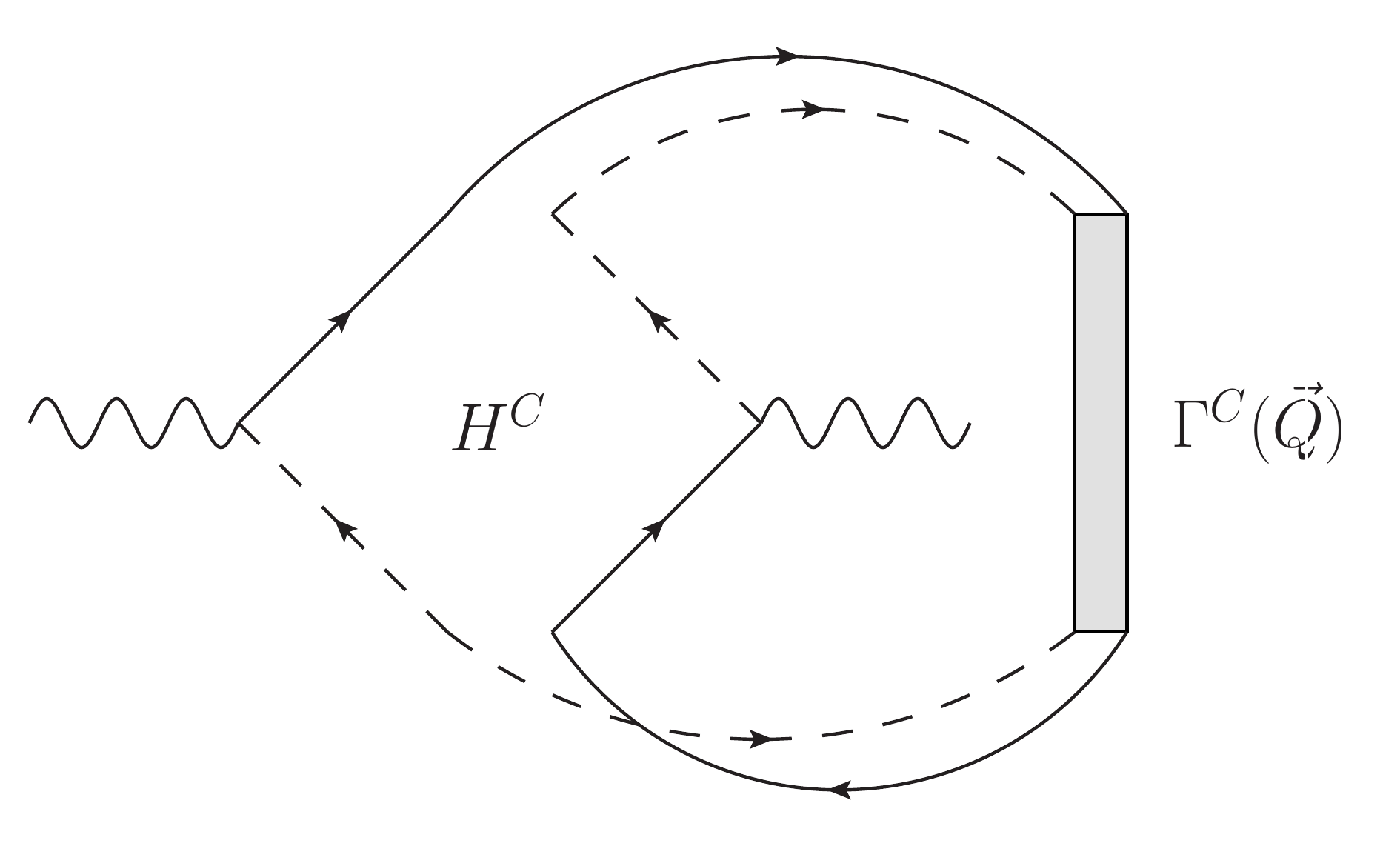}
\caption{
\label{fig:WAL_diagram}
Diagrammatic representation of the quantum correction to conductivity}
\end{figure}
\begin{figure} [!h]
\centering 
\includegraphics[width=13cm]{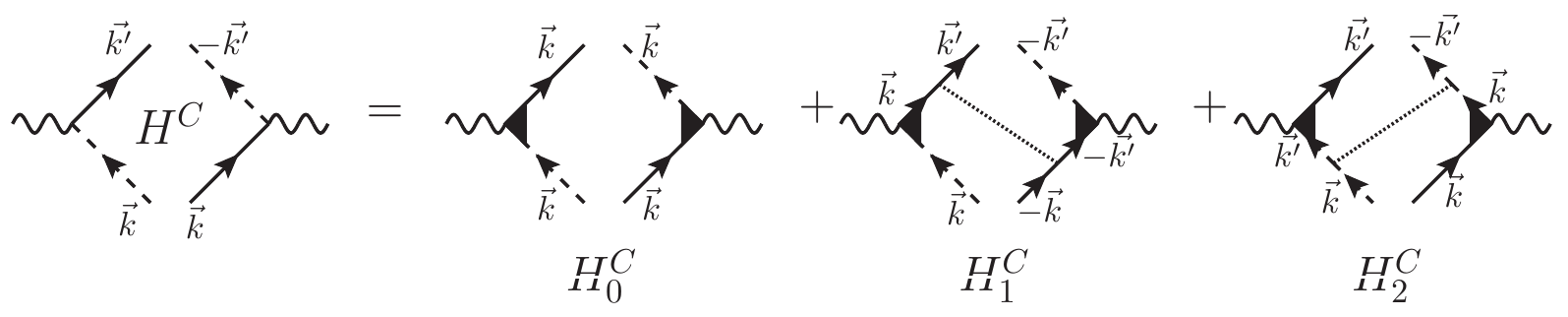}
\caption{
\label{fig:hikami_box}
Diagrammatic representation of the dressing of the Hikami box}
\end{figure}

The weak anti-localization correction is obtained by the contraction of a Cooperon propagator and a Hikami box, as represented diagrammatically on Fig.~\ref{fig:WAL_diagram}. This Hikami box is the sum of three different contributions represented 
in Fig.~\ref{fig:hikami_box}. We express the first of these contributions as 
\begin{equation}\label{eq:WAL-Hikami1}
\langle \delta \sigma_{0} \rangle = \frac{\hbar}{2\pi} \mathrm{Tr} 
\left[
G^{A}(\vec{k}) \Sigma^x G^{R}(\vec{k})  \Gamma^C (\vec{Q}) G^{R}(\vec{Q}-\vec{k})\Sigma^x G^{A}(\vec{Q}-\vec{k})
\right], 
\end{equation}
where we used the notations introduced in the article $\mathrm{Tr}$ for the trace over all the quantum numbers (spin and momenta), 
$\mathrm{tr}$ for the trace over the spin indices, 
 and $\int_{\vec{k}}$ for the trace over the momentum $\vec{k}$. 
 Special care has to be devoted to the order of the spin indices in these expressions.
In the expression Eq.(\ref{eq:WAL-Hikami1}), the $\vec{Q}$ integral is dominated by the small $\vec{Q}$ contribution originating from the diffusive modes of the Cooperon. 
This justifies {\it a posteriori} the projection on the single diffusive mode in (\ref{eq:Cooperon-propagator}). Focusing on the most dominant part of this expression, we can set $Q\to 0$ except in the Cooperon propagator, 
the Green's functions being regular in $\vec{k}$.
 Similarly, we can pull the renormalized vertices out of the integral and focus on the Hikami box :
\begin{eqnarray}
H^C_0 &=& 4
\int_{\vec{k}}
\left[
 G^{A}(\vec{k}) \sigma^{x} G^{R}(\vec{k})
\right] \otimes  \left[
G^{R}(-\vec{k}) \sigma^{x} G^{A}(-\vec{k})
\right]
\nonumber \\
& = &
  \rho (E_{F})
\left( \frac{ 2 \tau_{e}}{\hbar} \right)^{3}
\frac{\pi}{16}
\left[3  \sigma^{x}\otimes \sigma^{x} +  \sigma^{y} \otimes \sigma^{y} -4 ~\mathbb{1} \otimes \mathbb{1}
\right]. 
\end{eqnarray}
Similarly, the two remaining diagrams contributing to the Hikami box in Fig.~\ref{fig:hikami_box} are expressed as 
\begin{eqnarray}
H^C_1 &=&
4\gamma~
\int_{\vec{k}}
\int_{\vec{q}_{1}}    
\left[
G^{A}(\vec{k}) \sigma^{x} G^{R}(\vec{k}) G^{R}(-\vec{q}_{1}) 
\right]\otimes \left[
G^{R}(-\vec{k}) G^{R}(\vec{q}_{1})  \sigma^{x}  G^{A}(\vec{q}_{1})
\right]
 \nonumber \\
&=&
\frac{\pi}{16}  \rho(E_{F})  \left( \frac{ 2 \tau_{e}}{\hbar} \right)^{3}
\left[ \mathbb{1}\otimes  \mathbb{1}   - \sigma^{x} \otimes \sigma^{x} \right]
\end{eqnarray}
 and : 
\begin{eqnarray}
H^C_2 &=& 
4 \gamma~
\int_{\vec{k}}
\int_{\vec{q}_{1}}    
\left[
G^{A} (-\vec{q}_{1}) G^{A}(\vec{k}) \sigma^{x} G^{R}(\vec{k})  
\right]
\otimes 
\left[
G^{R}(\vec{q}_{1}) \sigma^{x} G^{A}(\vec{q}_{1}) G^{A}(-\vec{k})
\right]
\nonumber \\
&=&
\frac{\pi}{16}  \rho(E_{F})  \left( \frac{ 2 \tau_{e}}{\hbar} \right)^{3}
\left[ \mathbb{1}\otimes  \mathbb{1}   - \sigma^{x} \otimes \sigma^{x} \right]
\end{eqnarray}
Summing these three contributions we obtain : 
\begin{equation}
H^C  =  
  \rho (E_{F})
\left( \frac{ 2 \tau_{e}}{\hbar} \right)^{3}
\frac{\pi}{16}
\left[  \sigma^{x}\otimes \sigma^{x} +  \sigma^{y} \otimes \sigma^{y} -2 ~\mathbb{1} \otimes \mathbb{1}
\right]. 
\end{equation}
The resulting weak anti localization correction is obtained {\it via} the final contraction with a Cooperon propagator 
as shown on Fig.\ref{fig:WAL_diagram}, and leads to the expression: 
\begin{eqnarray}
\langle \delta \sigma \rangle    &= &
\frac{\hbar  \rho (E_{F}) (-ev_{F})^2 }{2\pi} 
\left( \frac{ 2 \tau_{e}}{\hbar} \right)^{3}
\frac{\pi}{16}  \mathrm{tr}
\left(  \sigma^{x}\otimes \sigma^{x} +  \sigma^{y} \otimes \sigma^{y} -2 ~\mathbb{1} \otimes \mathbb{1}
\right)
\int_{\vec{Q}}  \Gamma (\vec{Q})
 \nonumber \\
 &= &\left( \frac{e^2}{\pi \hbar} \right)  \int_{\vec{Q}}\frac{1}{Q^2}
\end{eqnarray}

\section{Universal Conductance Fluctuations for Dirac fermions}
\label{app:UCF}

\begin{figure} [!h]
\centering 
\includegraphics[width=0.7\textwidth]{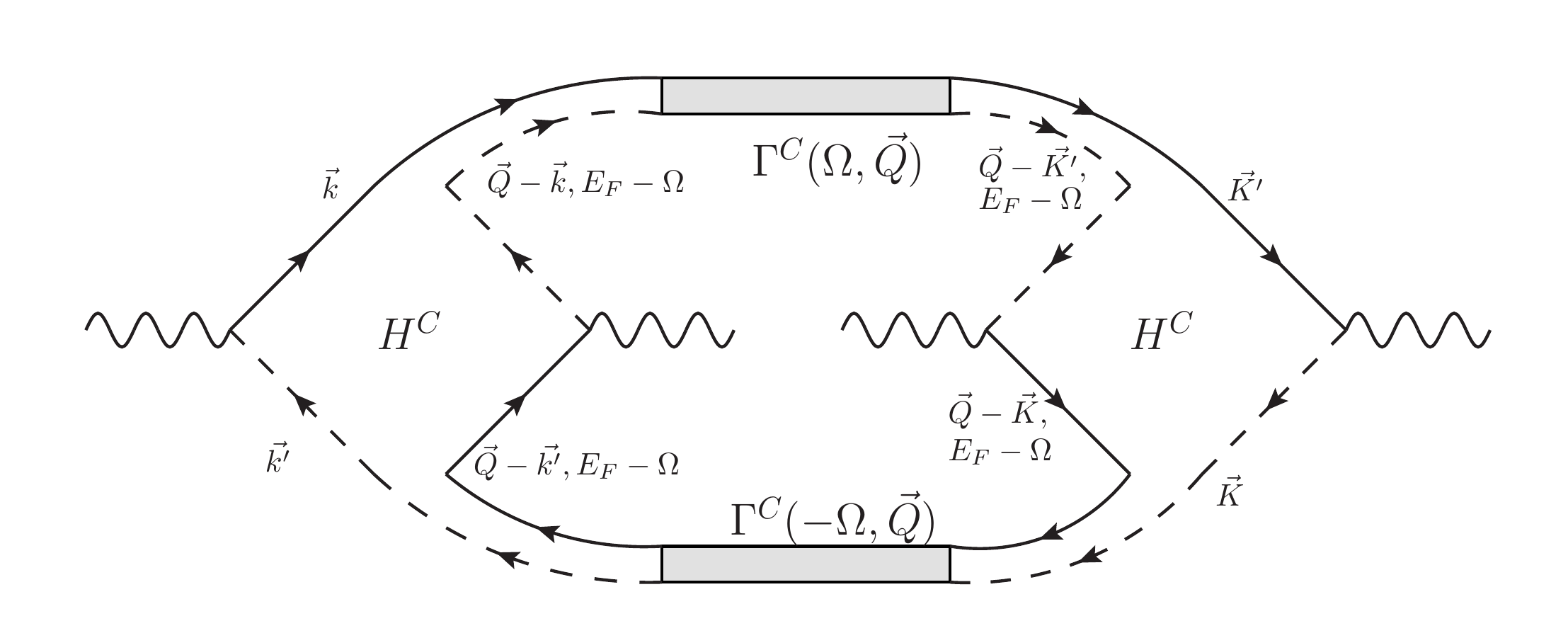}
\caption{Diagram for the conductance fluctuations with Cooperons}
\label{UCF_1_coop}
\end{figure}

\begin{figure} [!h]
\centering 
\includegraphics[width=0.7\textwidth]{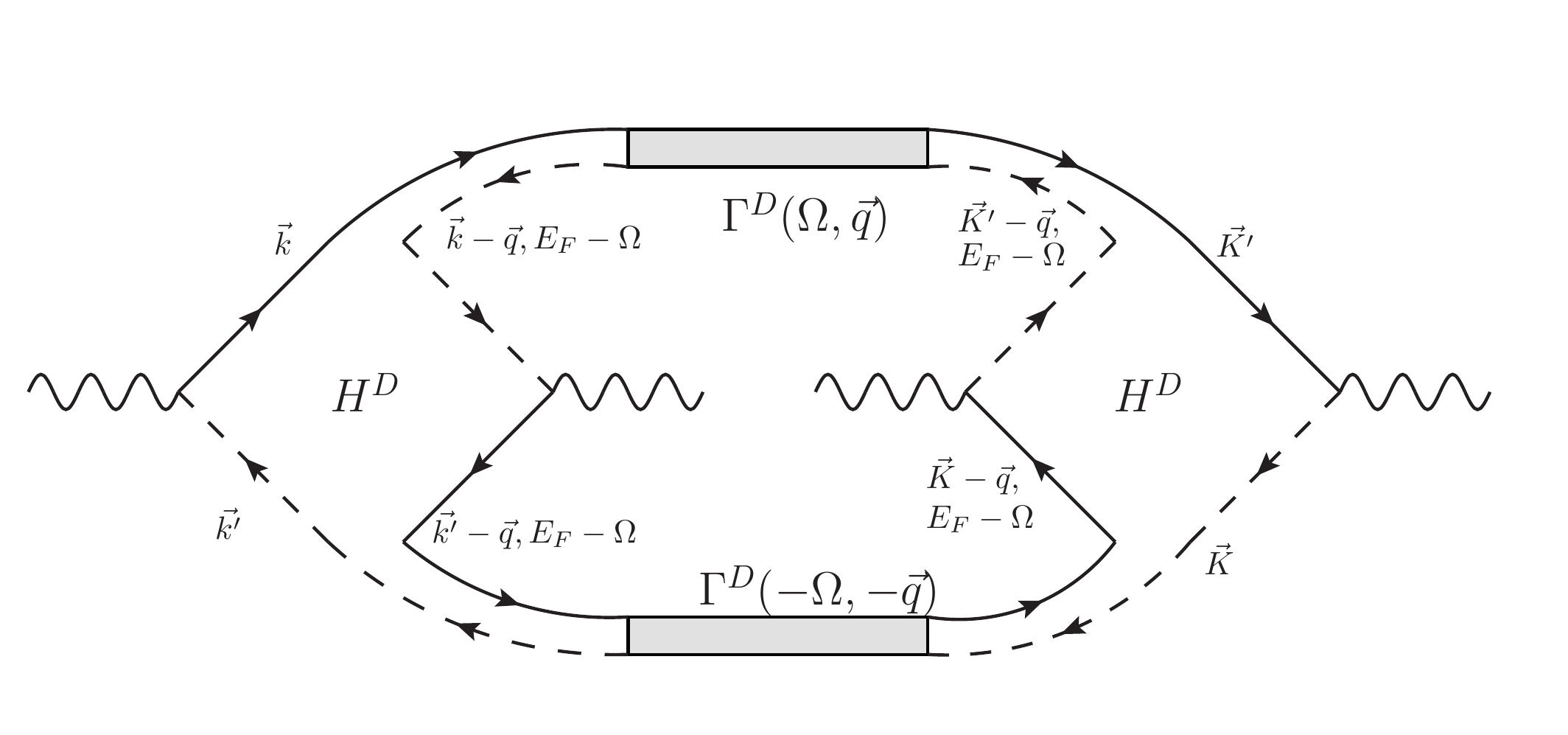}
\caption{Diagram for the conductance fluctuations with Diffusons.}
\label{UCF_1_diff}
\end{figure}

To derive the conductance fluctuations, we need to take into account two kinds of diagrams containing either Cooperons or 
 Diffusons. The Hikami box for Cooperons has been calculated previously. Proceeding similarly with the Diffuson instead of 
 Cooperon structure factor  we obtain the Hikami box for Diffusons :
\begin{equation}
H^D = 
  \rho (E_{F})
\left( \frac{ 2 \tau_{e}}{\hbar} \right)^{3}
\frac{\pi}{16}
\left[  2 ~\mathbb{1} \otimes \mathbb{1} + \sigma^{x}\otimes \sigma^{x} +  \sigma^{y} \otimes \sigma^{y} 
\right]. 
\end{equation}
We have already performed an integration over the momentum (arising from the Kubo formula) in these expressions for the 
 Hikami boxes. Hence we only need to plug a Diffuson (resp. Cooperon) structure factor between two $H^D$ (resp. $H^C$). 
Summing these two diagrams (Fig. \ref{UCF_1_diff} and Fig. \ref{UCF_1_coop}) we obtain  :
\begin{equation}\label{eq:UCF-contrib1}
\langle \delta \sigma^2_1 \rangle = 8 \left( \frac{e^2}{h} \right)^2 \  \frac{1}{V}\! \int_{\vec{q}}\frac{1}{q^4} . 
\end{equation}
The second part of the conductance fluctuations come from the diagrams represented in Fig. \ref{UCF_2} that we have not yet considered.  
They require the determination of two additional Hikami boxes (one for Diffusons and one for Cooperons) : 
\begin{figure} [!h]
\centering 
\includegraphics[width=\textwidth]{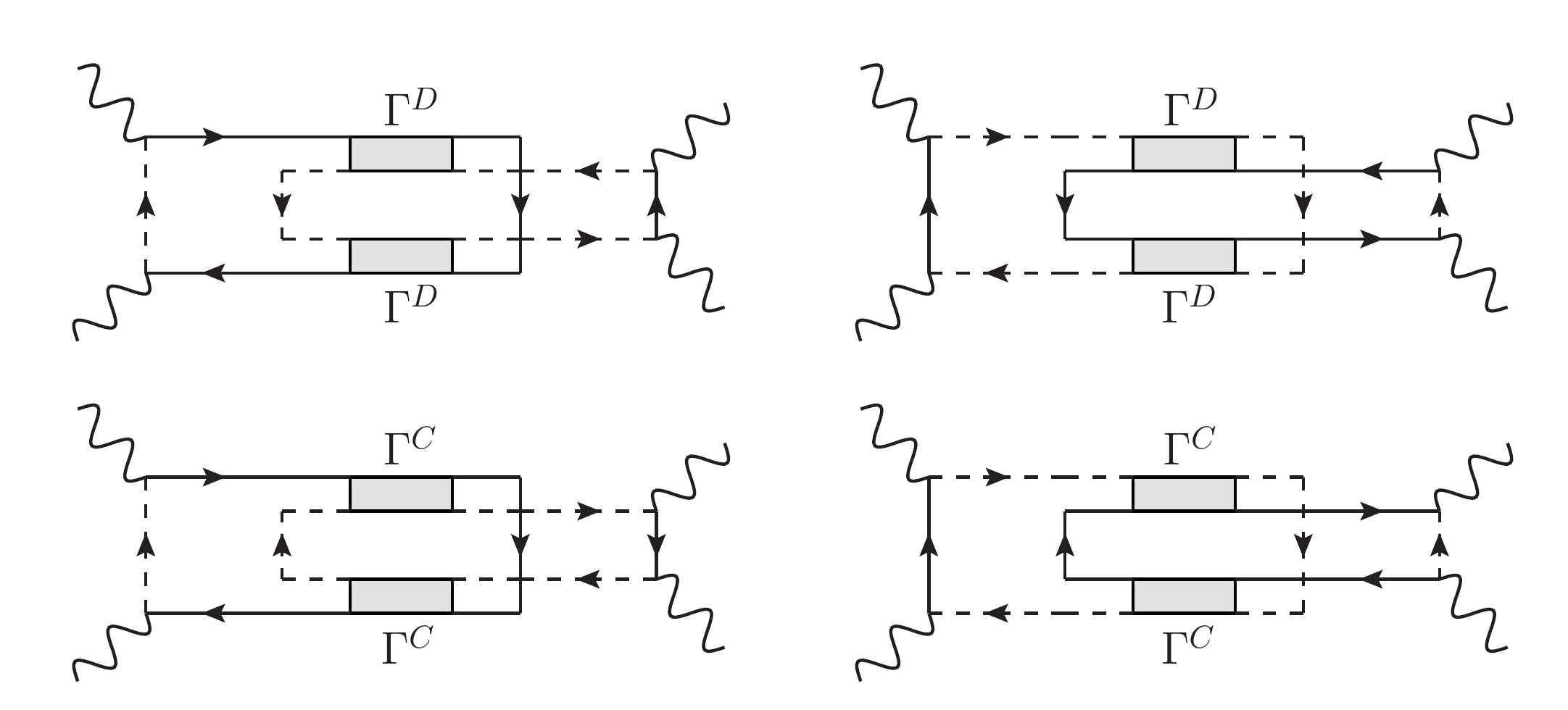}
\caption{Diagrams for the second contribution to conductance fluctuations}
\label{UCF_2}
\end{figure}
\begin{eqnarray}
\tilde{H}^D &= & 
  \rho (E_{F})
\left( \frac{ 2 \tau_{e}}{\hbar} \right)^{3}
\frac{\pi}{16}
\left[  \mathbb{1} \otimes \mathbb{1} + \sigma^{x}\otimes \sigma^{x} \right] \\
\tilde{H}^C& = & 
  \rho (E_{F})
\left( \frac{ 2 \tau_{e}}{\hbar} \right)^{3}
\frac{\pi}{16}
\left[  \mathbb{1} \otimes \mathbb{1}- \sigma^{x}\otimes \sigma^{x}  
\right].  
\end{eqnarray}
 The final results after contraction in spin space of these diagrams is 
\begin{equation}
\label{eq:UCF-contrib2}
\langle \delta \sigma^2_2 \rangle = 4 \left( \frac{e^2}{h} \right)^2 \  \frac{1}{V}\! \int_{\vec{q}}\frac{1}{q^4}. 
\end{equation}
Summing the two contributions (\ref{eq:UCF-contrib1}) and (\ref{eq:UCF-contrib2}), we finally get the result of Eq \ref{eq:UCF-Dirac} :
\begin{equation}
\langle \delta \sigma^2 \rangle = \langle \delta \sigma^2_1 \rangle+\langle \delta \sigma^2_2 \rangle = 12 \left( \frac{e^2}{h} \right)^2 \  \frac{1}{V}\! \int_{\vec{q}}\frac{1}{q^4}
\end{equation}

\section{Quantum correction for a warped Fermi surface}

 \ref{app:WAL} and \ref{app:UCF} show explicitly the derivation of the weak antilocalization correction and the conductance fluctuations for Dirac 
fermions. 
As expected, the corresponding results display no dependence in the only relevant parameter to characterize diffusion : 
 the diffusion constant.  These results are naturally expected to hold when taking into account the hexagonal warping term. 
 To explicitly show this independence, we  determine the value of the quantum correction to conductivity in the 
general case where the Fermi surface possesses the hexagonal deformation.

The first step is to obtain the new Hikami box $H^C$, the difficulty arising from the dependence of the current operator 
on the momentum
$j_{x}  = e \left( - v_{F} \sigma_x + \frac{3 \lambda}{ \hbar} \sigma_z (k_x^2-k_y^2) \right)$. 
Recalling that $\Sigma^{x} = j_{x} + j_{x}  P^{D} \Gamma^{D}$, we express the "naked" Hikami box as:
\begin{equation}
H^C_0 =  \int_{\vec{k}}
\left[
 G^{A}(\vec{k}) \Sigma^{x} G^{R}(\vec{k})
G^{R}(-\vec{k}) \Sigma^{x}  G^{A}(-\vec{k})
\right]. 
\end{equation}
We  perform this integral using polar coordinates, and integrate the radial part :
\begin{eqnarray}
H^C_0 & = & \frac{e^2 v_F^2 \tau_e^2}{4 \gamma \hbar} \frac{\tau_e}{\tau_e^{(0)}} \int \frac{d \theta}{2 \pi} \frac{A_+ B A_+ \otimes A_- B A_-}{1 + 12 b^2 \tilde{k}(\theta)^4 \cos^2(3 \theta)} \\
A_{\pm} & = & \mathbb{1} \pm \tilde{k}(\theta) \left(\cos \theta \sigma^x + \sin \theta \sigma^y \right) 
+ 2 b~ \tilde{k}^{3}(\theta) \cos(3 \theta) \sigma^z \\
B & = & \left(2 + \frac{\beta}{\alpha + \beta}\right)\sigma^x + 6b~ \tilde{k}^{2}(\theta) \cos(2 \theta) \sigma^z
\end{eqnarray}

The two "dressed" Hikami boxes, where an impurity line links two Green's functions can be calculated by the method used above, and we can write the result as $H^C = \frac{e^2 v_F^2 \tau_e^2}{4 \gamma \hbar} H(b)$. We need to close this Hikami box with a Cooperon structure factor $\Gamma^C = \frac{\gamma}{\tau_e} \frac{1}{D(b)Q^2} \frac{1}{4} \left(\mathbb{1} \otimes \mathbb{1} - \sigma^x \otimes \sigma^x  - \sigma^y \otimes \sigma^y -  \sigma^z \otimes \sigma^z \right)$. 
Performing 
the sum on the spin space $h(b) = \mathrm{tr} [
H(b) \frac{1}{4} \left( \mathbb{1} \otimes \mathbb{1} - \sigma^x \otimes \sigma^x  - \sigma^y \otimes \sigma^y -  \sigma^z \otimes \sigma^z \right)] $ we obtain  the quantum correction to the conductivity :
\begin{equation}
\langle \delta \sigma (b) \rangle = \frac{\hbar}{2 \pi} \mathrm{Tr} (H^C \Gamma^C) = \frac{e^2}{h} \frac{v_F^2 \tau_e h(b)}{2 D(b)} \int \frac{d \vec{q}}{q^2} .
\end{equation}
We found numerically  the correction $\frac{v_F^2 \tau_e h(b)}{2 D(b)}$ to be constant  equal to $1$ and independent of $b$.

\end{appendix}

\medskip

\bibliographystyle{unsrt}

 \end{document}